\documentclass[apjl]{emulateapj}
\usepackage{natbib}
\usepackage{times}
\usepackage{graphicx}
\usepackage{amsmath}

\newcommand{\zlim}{0.0342}
\newcommand{\kms}{{\rm km \, s}^{-1}}
\newcommand{\seclim}{-18.3}
\newcommand{\prilim}{-20.8}
\newcommand{\seclimh}{-17.5}
\newcommand{\prilimh}{-20.0}
\newcommand{\medpri}{-21.3}
\newcommand{\medsec}{-19.3}
\newcommand{\magh}{- 5 \log(h)}
\newcommand{\LCDM}{$\Lambda$CDM}
\newcommand{\kpch}{\,\,{\rm kpc}/h}

\newcommand{\vmax}{v_{\rm max}}
\newcommand{\MSII}{MS-II}
\newcommand{\msun}{M_{\odot}}
\newcommand{\hmpc}{h^{-1} {\rm Mpc}}
\newcommand{\pks}[1]{p_{\rm KS}=#1 \%}

\begin{document}

\title{Small-Scale Structure in the SDSS and \LCDM{}: Isolated $\sim L_*$ Galaxies with Bright Satellites}
\shorttitle{SSS in SDSS and \LCDM{}}

\author{Erik J. Tollerud\altaffilmark{1}, Michael Boylan-Kolchin\altaffilmark{1,2}, Elizabeth J. Barton\altaffilmark{1}, James S. Bullock \altaffilmark{1}, Christopher Q. Trinh\altaffilmark{3,1}}
\altaffiltext{1}{Center for Cosmology, Department of Physics and Astronomy, The University of California Irvine, Irvine, CA, 92697, USA; etolleru@uci.edu, ebarton@uci.edu, m.bk@uci.edu, bullock@uci.edu}
\altaffiltext{2}{Max-Planck-Institut f¨ur Astrophysik, Karl-Schwarzschild-Str. 1, 85748 Garching, Germany}
\altaffiltext{3}{Sydney Institute for Astronomy, School of Physics, University of Sydney, NSW 2006, Australia; ctrinh@physics.usyd.edu.au}

\begin{abstract}

We use a volume-limited spectroscopic sample of isolated galaxies in the Sloan Digital Sky Survey (SDSS) to investigate the frequency and radial distribution of luminous ($M_r \lesssim -18.3$) satellites like the Large Magellanic Cloud (LMC) around $\sim L_*$ Milky Way analogs and compare our results object-by-object to \LCDM{} predictions based on abundance matching in simulations.
We show that $12 \%$   of Milky Way-like galaxies host an LMC-like satellite within 75 kpc (projected), and $42 \%$ within 250 kpc (projected).  This implies $\sim 10 \%$ have a satellite within the distance of the LMC, and $\sim 40 \%$ of $L_*$ galaxies  host a bright satellite within the virialized extent of their dark matter halos.  Remarkably, the simulation reproduces the observed frequency, radial dependence, velocity distribution, and luminosity function of observed secondaries exceptionally well, suggesting that \LCDM{} provides an accurate reproduction of the observed Universe to galaxies as faint as $L\sim10^9 L_{\odot}$ on $\sim 50$ kpc scales.      When stacked, the observed projected pairwise velocity dispersion of these satellites is $\sigma \simeq \; 160 ~ \kms$, in agreement with abundance-matching expectations for their host halo masses.  Finally, bright satellites around $L_*$ primaries are significantly \emph{redder} than typical galaxies in their luminosity range, indicating that environmental quenching is operating within galaxy-size dark matter halos that typically contain only a single bright satellite.   This redness trend is in stark contrast to the Milky Way's LMC, which is unusually blue even for a field galaxy.  We suggest that the LMC's discrepant color might be further evidence that it is undergoing a triggered star-formation event upon first infall.   
 \end{abstract}

\section{Introduction}

\label{sec:intro}
In the \LCDM{} concordance cosmology, dark matter substructure is ubiquitous due to the continuation of the power spectrum to very small scales.  As a result, substructure provides valuable constraints both on the nature of dark matter and its effects on galaxy formation.  The missing satellites problem---the observation that $\sim L_*$ galaxies like the Milky Way (MW) have too few satellites for the predicted amount of dark matter substructure \citep{kl99msp,moo99msp}---has been discussed in depth for faint satellites.  Proposed solutions include steep slopes or truncations in galaxy formation \citep[e.g.][and references therein]{BKW,strigari09nat,bovill09reion,kop09,krav10rev}, as well as a significant component of observational bias \citep{toll08,walsh09,bull10stealth}.  

In contrast, less is known about more luminous satellites of $\sim L_*$ hosts, e.g. satellites like the Large Magellanic Cloud (LMC) or Small Magellanic Cloud (SMC).  These luminosity ranges are instructive to compare to \LCDM{}, as the substructure is significant for $\sim L_*$ hosts at these scales, while the satellites are still bright enough to be detected at cosmological distances.  Thus, unlike faint Galactic satellites, these brighter satellites can be studied outside of the Local Group (LG) with statistically large samples \citep{wgb06,woodsgeller07}.  Studying the small-scale structure (the halo substructure at cosmological distances) of isolated galaxy halos therefore provides a valuable test for \LCDM{}. 

In \LCDM{}, LMC-sized satellites may be rare around MW-like hosts \citep[e.g.,][]{benson02,stri07,kop09}.   \citet{bk09noplace} considered directly how common MW/LMC and MW/LMC/SMC systems are in a very high-resolution \LCDM{} simulation (the same discussed in \S \ref{sec:sims} below), showing they are potentially quite uncommon ($3-25 \% $), depending on the assumed MW mass.  \citet{busha10mccount} performed a similar analysis on a different simulation, with similar results\footnote{While these results, and ours, may show differences at a factor of two level, we show below that this is primarily due to the exact choice of definition for ``MW-like'' or ``LMC-like'' and thus all are in accord when compared appropriately.}. 

Additionally, there are observational hints that LMCs are unusual around MW-like hosts \citep{james08ha,james10mcs,liu10,vdb10mcs}.  Despite this possibility, models are sometimes tuned to produce the correct number of bright satellites like the MCs, or are advanced because they naturally do so \citep[e.g.][]{lib07,kop09,krav10rev}.  Thus, an understanding of the abundance and distribution of luminous satellites around isolated hosts constrains models of galaxy formation, and helps place the Local Group itself in a clearer cosmological context.

In this study, we present a set of spectroscopically selected galaxy pairs in the Sloan Digital Sky Survey (SDSS) composed of an isolated $\sim L_*$ primary and an $\sim L_{\rm LMC}$ secondary. We statistically connect these systems to \LCDM{} dark matter halos selected identically from simulations.  While we would ideally include SMC-like satellites, the SDSS spectroscopic survey is not deep enough to provide a statistically significant sample of SMC analogs. We thus probe the clustering properties of the lowest-luminosity satellites possible with SDSS spectroscopy, and thereby provide a robust test of \LCDM{} for the smallest subhalos and scales that can be probed with a statistical significant sample.  Given the similarity of our systems to the MW/LMC group, this approach also allows us to address how common such systems are in the local universe.

Throughout this paper we use the term ``primary'' and ``secondary'' to indicate the larger and smaller galaxies in our projected pair samples, for which the pair may not actually be physically associated.  In contrast, we use the terms ``host'' and ``satellite'' to indicate pairs for which the smaller galaxy is definitely within the larger's halo (and hence only applies for halos in the simulation, where full 3D information is available).

This paper is organized as follows: In \S \ref{sec:obs} we describe how we construct our observational sample, in \S \ref{sec:sims} we describe the simulation we use to interpret this sample, in \S \ref{sec:sampanl} we analyze the sample and compare to the expectations of \LCDM{}, in \S \ref{sec:lmc} we discuss the MW/LMC system in the context of this sample, in \S \ref{sec:comparison} we compare our results to recent related works, and in \S \ref{sec:conc} we review our primary results.

\section{Observational Data}
\label{sec:obs}

We begin by describing the SDSS data sample chosen to identify \emph{potential} host/nearest satellite pairs (i.e., primaries and secondaries) in a volume-limited sample.  We use the NYU Value-Added Galactic Catalog \citep[VAGC][]{blanton05VAGC} based on SDSS DR7 \citep{DR7} with the improved photometric calibrations \citep{SDSSphot}.  From this catalog, we examine the main galaxy sample, using K-corrected \citet{petrosian} magnitudes with $z=0$ band shift. We use the $r$-band, which typically has the highest signal-to-noise ratio of the SDSS bands. Where relevant, we assume WMAP7 \citep{WMAP7} cosmological parameters with $h \equiv H_0/100 = 0.704$.

We apply the following set of selection criteria in order to construct a clean sample of primaries and secondaries.  As  summarized in Table \ref{tab:samps}, we also construct samples using subsets of these criteria in order to test our method and to estimate completeness in our spectroscopic sample.

\begin{enumerate}

\item $M_{r,\rm pri} \magh < \prilimh$  (i.e., $M_{r,\rm pri} < \prilim$) for primaries, \label{crit:prilum}
\item $\prilimh < M_{r,\rm sec} \magh < \seclimh$ ($\prilim < M_{r,\rm sec}  < \seclim$) for secondaries, \label{crit:seclum}
\item $z_{\rm pri} < \zlim$, which ensures the sample is complete  based on SDSS spectroscopic survey completeness limits for $M_r \magh < \seclimh$ ($M_r < \seclim$).  \label{crit:vol}
\item The primary must have a measured redshift. \label{crit:priz}
\item The primary must be isolated: at most one other galaxy in the primary magnitude range within  $250 < d_{\rm proj} < 700 \kpch$ ($355 < d_{\rm proj} < 995$ kpc) and with redshifts such that $\Delta v_{\rm pri}<1000$ $\kms$.  There must be none within $250 \kpch$ ($355$ kpc). (see \S \ref{sec:iso}) \label{crit:iso}
\item The secondary must be within $250 \kpch$ ($355$ kpc) projected and be the nearest (projected) galaxy in the secondary magnitude range. \label{crit:closest}
\item The secondary must have a measured redshift. \label{crit:secz}
\item $\Delta v \equiv c|z_{\rm pri}-z_{\rm sec}| < 500$ $\kms$. \label{crit:dv}

\end{enumerate}

 \begin{deluxetable*}{cccccccc}
 \tablecolumns{8}
 \tablecaption{Summary of catalog samples.}
 \tablehead{
   \colhead{Sample Name \tablenotemark{1}} &
   \colhead{Mag/Vol Limit \tablenotemark{2}}&
   \colhead{Pri/Host Present \tablenotemark{3}} &
   \colhead{Sec/Sat Present \tablenotemark{4}} &   \colhead{Sec Has Redshift \tablenotemark{5}} &   \colhead{$\Delta v$ Cut \tablenotemark{6}} &   \colhead{Obs or Sim \tablenotemark{7}} &
   \colhead{N \tablenotemark{8}}
 }
  \startdata
 Primaries                   & y & y & y or n & N/A    & N/A    & Obs & 1075 \\ 
 Hosts                       & y & y & y or n & N/A    & N/A    & Sim & 373477 \\ 
 Clean                       & y & y & y      & y      & y      & Obs & 467 \\
 Simulated Clean             & y & y & y      & N/A    & y      & Sim & 149906 \\
 Any $z$                     & y & y & y      & y      & y or n & Obs & 584 \\
 No $z$                      & y & y & y      & n      & N/A    & Obs & 189 \\
 Full                        & y & y & y      & y or n & N/A    & Obs & 773 \\
 Control                     & y & n & y      & y      & N/A    & Obs & 24298 
 \enddata
 
 \tablenotetext{1}{Name of the sample.}
 \tablenotetext{2}{Whether or not magnitude and volume limits are applied (Criteria \ref{crit:prilum}-\ref{crit:priz}).}
 \tablenotetext{3}{Whether or not a primary (host in simulations) is present.}
 \tablenotetext{4}{Whether or not a secondary (satellite in simulations) is present (Criterion \ref{crit:closest}).}
 \tablenotetext{5}{Whether or not a secondary must have a redshift (Criterion \ref{crit:secz}).}
 \tablenotetext{6}{Whether or not a secondary (satellite in simulations) must be within 500 $\kms$ of its primary/host (Criterion \ref{crit:dv}).}
 \tablenotetext{7}{Specifies if the data are from SDSS/VAGC galaxy catalog or the \MSII{} mock galaxy/halo catalogs.}
 \tablenotetext{8}{Number of objects in the sample.}
 
 \label{tab:samps}
\end{deluxetable*}

We now consider these criteria and their motivation in more detail.  We only consider galaxies with measured redshifts as potential primaries.  With these redshifts, we select a volume-limited sample of galaxies, with the maximum redshift set such that all primaries and secondaries  are within the completeness limit of the SDSS spectroscopic survey ($r<17.77$).  From these objects, we identify all galaxies more luminous than our primary/secondary boundary ($M_r<\prilim$) as primaries.  This sample has a median redshift of $z_{\rm med}=0.028$, near the upper limit due to the larger volume at higher redshifts.

Without further selection, many of these primaries are in groups or clusters.   In such dense environments, secondaries with even small projected separations from the primary are often not true satellites (see \S \ref{sec:sims}).  We therefore apply additional criteria to the primaries to ensure we are selecting relatively isolated primaries. Specifically, following \citet{barton07}, for each primary we identify all other primaries within a projected distance of 1 Mpc for which the redshift difference between the two satisfies $\Delta v_{\rm pri} \equiv c | z_1-z_2| < 1000 \; \kms$.  An isolated primary is then defined as having at most one primary within this distance, and if such a companion is present, it cannot be within $355$ kpc.  As estimated using cosmological simulations in \S \ref{sec:iso}, this criterion results in $\sim 90 \%$ of the secondaries being true satellites, or a contamination of $\sim 10 \%$, while still maintaining a large sample.

\begin{figure}[t!]
\epsscale{1.15}
\plotone{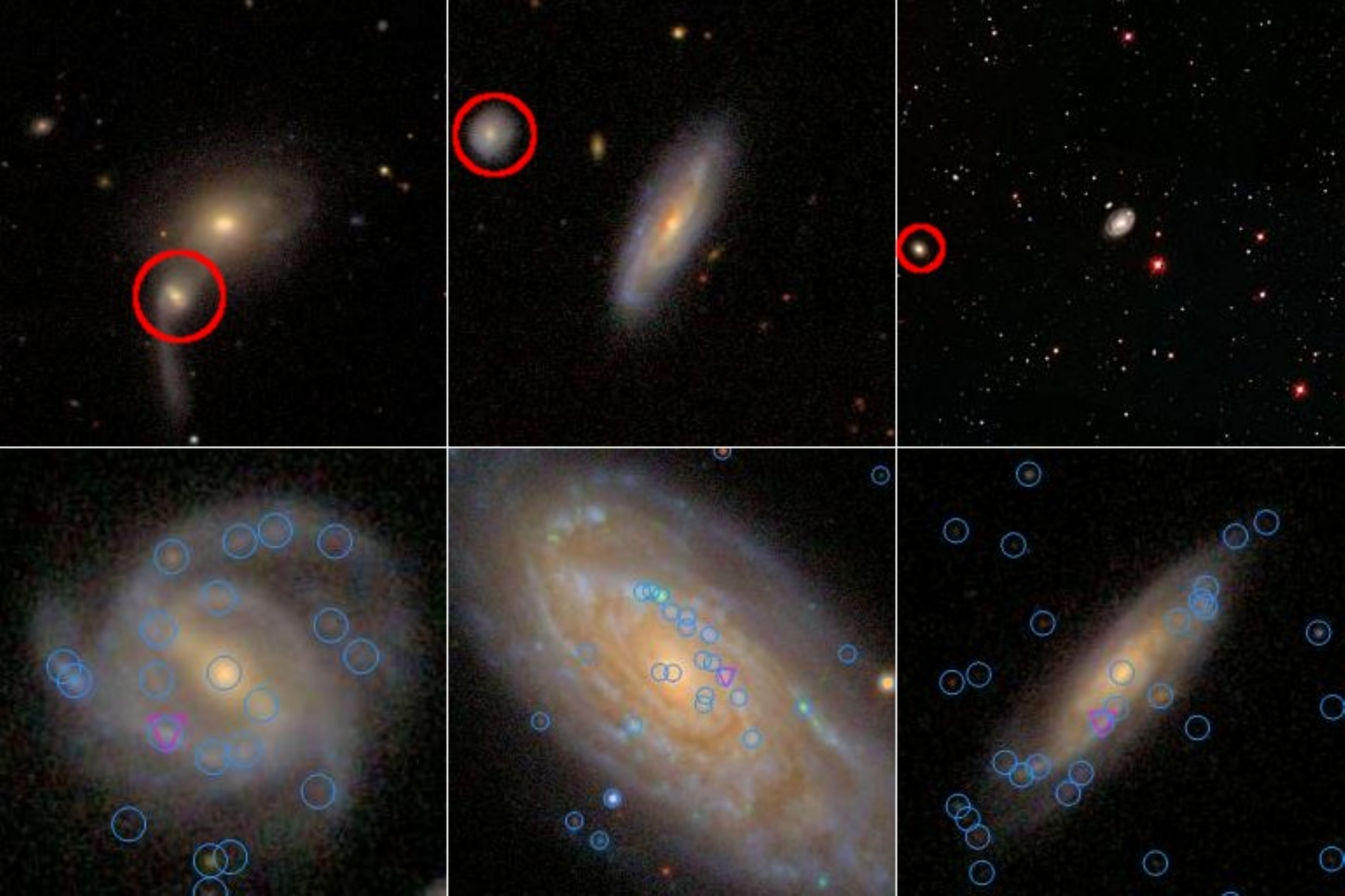}
\caption{Examples of SDSS primary/secondary pairs in the clean sample (upper) and false pairs (lower).  Secondaries identified by our criteria (see text) are marked with red circles (upper panels) or magenta triangles (lower panels).  The upper three  are all in the clean sample (have redshifts close to the primary) and span a range of projected separations.  For the lower three images, blue circles are SDSS pipeline photometric objects, clearly showing the identification of HII regions as  photometric objects.  For these same lower three, the secondaries are clearly HII regions in the primary (or satellites that are indistinguishable from HII regions).  We visually identify and remove all pairs of this kind from our sample. }
\label{fig:hiiex}
\end{figure}

With a sample of isolated primaries in hand, we then identify the \emph{nearest} galaxy fainter than this boundary, but more luminous than the limit for secondaries ($\prilim<M_r<\seclim$), and label this the secondary if it is within $355$ kpc projected.  For this step we assume that the redshift of the secondary is that of the host to determine absolute magnitudes and physical distances.  This defines our ``Full'' sample of primary/secondary galaxies.  To clean this sample of foreground/background galaxies that only appear associated with the primary in projection, we filter this sample such that the redshift of the secondary be measured (the ``any $z$'' sample) and also require the secondary's redshift to be within $500 \, \kms$ of the primary.  This defines our main sample of primary/secondary pairs (hereafter referred to as the ``clean'' sample), three of which we show in the upper panel of Figure \ref{fig:hiiex}.  The remaining samples are determined by not including various steps of this procedure, and are specified in Table \ref{tab:samps}.

One final filtering step is necessary due to imperfections in the SDSS photometric pipeline.  For nearby bright galaxies (a significant portion of our primary sample), individual HII regions in a star forming disk are sometimes classified as separate objects in the SDSS pipeline, as shown in the lower panel of Figure \ref{fig:hiiex}.  Because these HII regions are extended sources and have redshifts matching the host, they are included in all secondary samples, including the clean sample.  Fortunately, this effect is only significant out to separations of $\sim 10$ kpc, so there are a small enough number of possible such pairs ($\sim50$) that they can be removed by visual examination.  All samples below have been cleaned of these HII regions.  This process is not ideal, as it may remove faint satellites that truly are present, but are superimposed over a disk and thus appear indistinguishable from an HII region.  Alternatively, some secondaries may be hidden from view by being behind the primary.  Thus, removing apparent HII regions may result in under-counting satellites at small projected separations.  Fortunately, this is only of consequence for $d_{\rm proj} \lesssim 10$ kpc, representing only $\sim 0.3 \%$ of a typical primary's dark matter halo volume. Hence, these under-counting effects are not significant relative to Poisson errors in the discussion below.

\begin{figure}[t!]
\epsscale{1.27}
\plotone{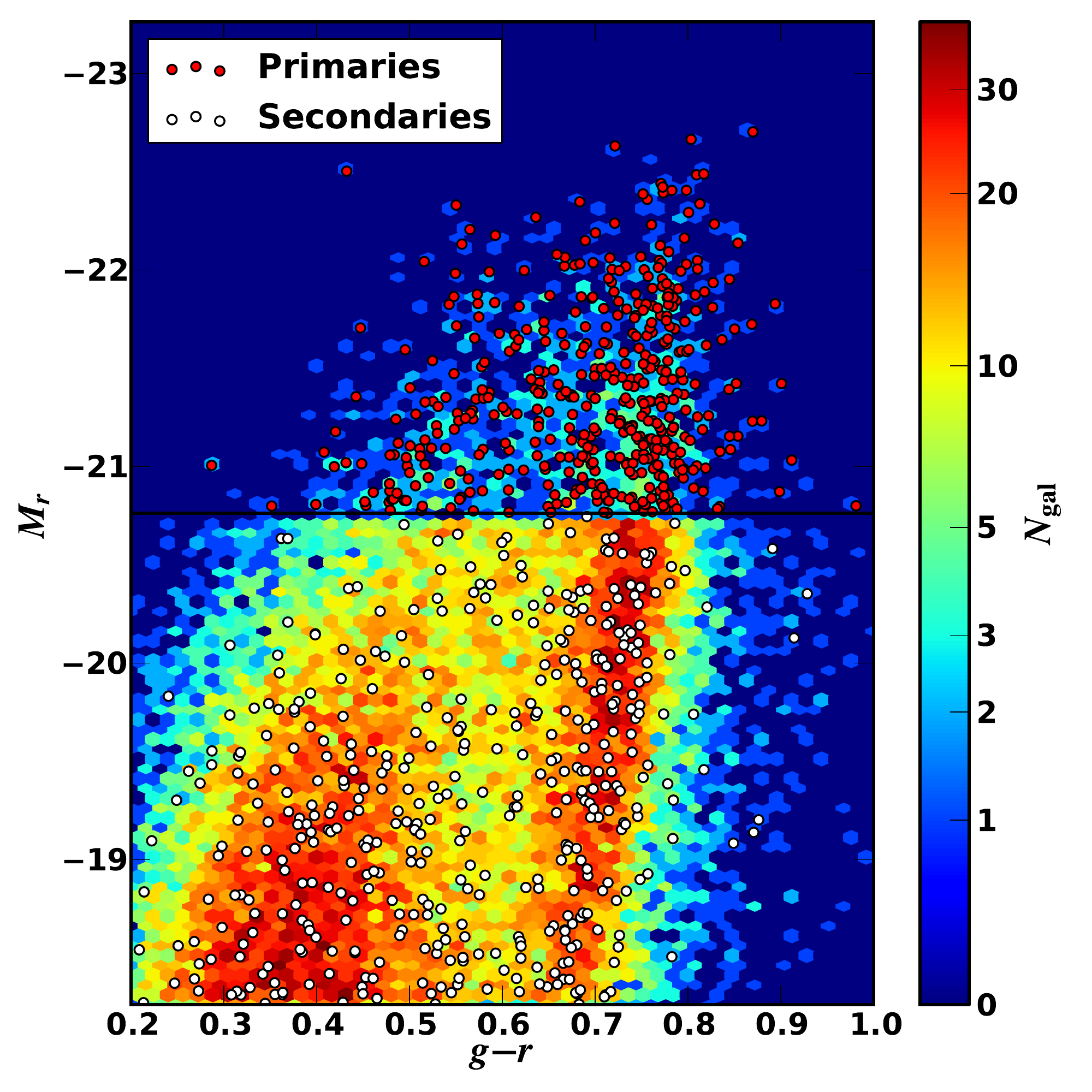}
\caption{Color-magnitude diagram ($g-r$ and $M_r$) for the SDSS data set.  Over-plotted circles are the clean sample: primaries with secondaries are colored red (above the horizontal black line), and their secondaries (below) are white.  The underlying hex-binned plot is a Hess-like diagram of the number of galaxies in each bin.  Above the separation line, it shows the counts of all objects in the primary sample (isolated and inside the volume limit).  Below, the bins are counts of objects in the Control sample (secondaries in the volume limit).  Thus, the discontinuity in the bins at the line is due to application of the isolation criterion above and not below.}
\label{fig:matchcmd}
\end{figure}

The clean selection criteria define a sample of galaxy pairs that are isolated from other luminous galaxies, and composed of a $\sim L_*$ primary with a nearest satellite with $L \sim L_{\rm LMC}$.  We show the $g-r$, $M_r$ color-magnitude diagram (CMD) as Figure \ref{fig:matchcmd}, with the clean sample represented by the red (primary) and white (secondary) points plotted over all isolated primaries (above the black line) and all galaxies with secondary-like magnitudes (below) in the volume-limited sample.  We note here that because the luminosity function climbs sharply for fainter galaxies, the typical luminosity of the clean sample is close to the faint cutoff for these samples: for the isolated primaries, the median absolute magnitude is $M_r=\medpri$ (very close to $M_*$ from \citealt{blanton03lf}), while for the secondaries, it is $M_r=\medsec$.   It is important to also note that the secondary sample is not complete because the SDSS spectroscopic survey is not complete.  The two main sources of systematic incompleteness are due to two fiber collision effects.  The first is collision between the primary and secondary, but this does not significantly bias our sample because the typical primary-secondary separation is larger than the on-sky fiber diameter for most of our objects.  The second effect concerns the case where a potential satellite is near a bright background cluster, and hence is not available to be targeted for a spectrum in the SDSS sample.  This effect is discussed further in \S \ref{sec:sampanl}.  

\interfootnotelinepenalty=10000

The selection criteria discussed above define a sample roughly analogous to the MW/LMC system \footnote{While the presence of M31 may seem to imply the MW is not ``isolated'', the LG is in fact most likely composed of two isolated halos following the definition of isolation adopted here. Thus, if viewed from far away with typical redshift errors, the LG would be statistically impossible to distinguish from two more widely separated galaxies for most orientations.}. While there are 467 MW/LMC pairs in the clean sample, we find only 7 objects if we instead use a volume limit appropriate for SMC-luminosity objects ($\sim 2$ mag fainter than the LMC).  As a result, we only consider MW/LMC-like pairs for the remainder of this work.  We defer further discussion specific to the LG to \S \ref{sec:lmc}. 

\interfootnotelinepenalty=100

\section{Catalogs from the Millennium II Simulation}
\label{sec:sims}

While the observational sample described in the previous sections is designed to produce a sample of isolated $L_*$ hosts with LMC-like satellites, the purity and completeness are not a priori known.  Hence, we compare a high-resolution cosmological simulation to our SDSS sample and search for host/satellite pairs, ``observing''  the simulation in the same way we observe the SDSS data \citep{barton07}.  We primarily use this simulated catalog to make direct comparison with the SDSS sample, thus factoring out as many selection effects as possible in testing the theory.  Moreover, these simulated observations can be used to validate our observational sample selection by comparing the perfect position/velocity information in the simulation to our observational selection criteria.  This step is crucial for informing the observations, as it would otherwise be nearly impossible to interpret the observations due to the influence of projection effects and peculiar velocities.

We use the high-resolution Millennium II simulation (\MSII{}) of \citet{bk09millii} as our source of dark matter halos with the assumption that all massive dark matter halos host galaxies.  The simulation resolves a volume of $137^3 \,{\rm Mpc}^3$, a nearly identical volume to that probed by our SDSS sample ($z<\zlim$), evolved from $z=127$ to 0.  The background cosmology is:
\begin{eqnarray}
 \label{eq:cosmo_params}
 & & \Omega_{\rm tot} = \! 1.0, \; \Omega_m = \!0.25, \; \Omega_b=0.045, \;
 \Omega_{\Lambda}=0.75, \nonumber \\
 & & h = 0.73, \;\sigma_8=0.9, \; n_s=1\,,
\end{eqnarray}
where $h$ is the reduced Hubble constant at redshift zero, $\sigma_8$ is the rms amplitude of linear mass fluctuations in
$8 \,\hmpc$ spheres at $z=0$, and $n_s$ is the spectral index of the primordial
power spectrum.  The force resolution (Plummer-equivalent softening) in the
simulation is $1.4 \, {\rm kpc}$ (comoving), and the mass resolution is
$9.43\times 10^{6}\,\msun$.  Dark matter structures are identified by first
using a Friends-of-Friends (FOF) algorithm \citep{davis85} to identify
distinct halos, then applying the {\tt SUBFIND} algorithm \citep{springel01} to these halos to find substructure.  Subhalos are
also linked across simulation outputs in merger trees.  The combination of these
analyses enables us to reliably track redshift zero properties and histories of
all subhalos with maximum circular velocities $\vmax$ in excess of $30 \, \kms$.  
For further details of the simulation, halo catalogs, and merger trees, see \citet{bk09millii}.

\subsection{Abundance Matching}
\label{sec:abundmatch}

To connect our SDSS galaxies to \MSII{} halos, we require a mapping from galaxy luminosity to halo maximum circular velocity $\vmax$. Here we use the well-established subhalo abundance matching technique to
do this mapping \citep[e.g.][]{krav04hod,CW09,moster10,guo10}.  The $v_{\rm max}$ values from the simulation are
computed for all subhalos at $z=0$ as either: (1) the current value of $v_{\rm
max}$, if the subhalo is dominant in its FOF group, or; (2) the value of $v_{\rm
max}$ the subhalo had at the time its mass was at a maximum, if the subhalo is
non-dominant.  Luminosities are then matched to maximum circular velocities by
setting $n(>L) = n(>v_{\rm max})$.  

Matching the halo catalogs from the MS-II to
the galaxy luminosity function of \citet{blanton03lf}, we find our lower limit for selection of host halos is $\vmax > 166.5 \, \kms$ from $M_{r,\rm host}< \prilim$, and the constraint for satellite (sub)halos is $166.5 > \vmax > 94.8$ $\kms$, from $M_{r,\rm host} < \seclim$. Additionally, given the median observed $M_r$  values of $\medpri$ and $\medsec$ for the primary and secondary samples (see \S \ref{sec:obs}), with abundance matching we can estimate the median (sub)halo $\vmax$ for our galaxies.  For primaries and secondaries, this is 211 km/s and 112 km/s, respectively. These are both within 1 km/s of the median $\vmax$ in the Simulated Clean sample, implying our abundance matching results are consistent between our subsamples.

We note that by using the parameters for the selection criteria directly from the mock catalog, we are ignoring the effects of observational errors in the luminosity function. To quantify the effects of these errors, we perform 1000 Monte Carlo simulations of the luminosity function by sampling each bin from the luminosity function of \citet{blanton05dwarflf} assuming Gaussian errors.  For each of these, we compute the implied $\vmax$ using the abundance-matching scheme described above. For both the primary and secondary limits, the scatter about the median $\vmax$ is $ \approx 0.3\%$.  

Beyond the observational uncertainty in the luminosity function, there is an intrinsic galaxy-to-galaxy scatter in abundance matching due to stochastic effects in galaxy formation.   This scatter has been quantified in \citet{beh10}, but their analysis is for matching halo mass to stellar mass, where the systematic errors in estimating stellar mass are the dominant effect.  Because we use luminosity instead of stellar mass, observational systematics are less important, but the results are potentially color-dependent. Additionally, even random scatter in abundance matching will result in a systematic shift of our $\vmax$ limits inferred from our $M_r$ limits.  This is because the mass and luminosity functions are both rising to the lower mass/faint end, so scatter in abundance matching will shift more galaxies from below to above the limit than vice versa.  This has the effect of systematically biasing the $\vmax$ limit for a given $M_r$.  

To estimate the magnitude of this bias in the $\vmax$ limit for a given $M_r$  for our data set, we use the results of the cosmological smoothed particle hydrodynamics (SPH) simulations of \citet{simha10sham}.  They find that the $1 \sigma$ log-scatter of the ratio of the luminosity of a galaxy to the luminosity assigned by abundance matching is $0.15$ dex. For each of the \MSII{} halos, we assign a median luminosity by abundance matching to the \citet{blanton05dwarflf} luminosity function.  We then draw 1000 samples from a lognormal distribution about this median with $\sigma=0.15$ dex.  We use this ensemble of halo luminosities to produce 1000 simulated luminosity functions, and abundance match them to the halo $\vmax$ function.  Finally, for each abundance matching mapping, we determine the resulting $\vmax$ limits for our primary and secondary cutoffs of $M_r < \prilim$ and $M_r < \seclim$.  We find that the resulting $\vmax$ limits are indeed systematically biased low, but by $< 1 \%$, which is much smaller than the scatter due to different lines-of-sight in \MSII{} (an estimate of cosmic variance).  To produce scatter comparable to this cosmic variance effect, the scatter in luminosity must be a factor of $\sim 10$, an implausibly large scatter based on the results of \citet{beh10}.

\subsection{Mock Galaxy Catalogs}
\label{sec:mockgal}

With the delineation between hosts and satellites in the simulation determined
(hosts: $\vmax > 166.5$ $\kms$; satellites: $\vmax > 94.8$ $\kms$), we proceed to
create mock observations of the \MSII{} in a fashion similar to \citet{barton07}.  As described in \S \ref{sec:obs}, the key elements of the observational sample are a maximum distance for the volume limit (and associated minimum luminosity), a lower luminosity limit for primaries, a lower luminosity limit for secondaries, isolation criteria for the primaries, and a cut on the difference in line-of-sight velocity between the primary and secondary.  Table \ref{tab:samps} specifies the exact criteria for the specific samples.  Here we describe specifically how this sample is extracted from the \MSII{} catalogs.

A random position within the box is chosen for the ``observer'', and a random
rotation to the simulation volume is applied, placing the observer at the origin
with the simulation volume in the positive (x, y, z) octant.  Line-of-sight
distances to all potential hosts in the simulation volume are computed, and for
each of these, transverse distances to all other objects in the volume are
calculated as well.  All hosts within a transverse distance of 1 Mpc and with a
line-of-sight velocity separation less than 1000 $\kms$ of each host are then
saved, as is the satellite with the smallest transverse distance.  As in for
observational data, we require (1) zero (at most one) additional hosts within
355 kpc (between 355 and 1000 kpc) of each host; (2) $z_{\rm pri}, \,z_{\rm
sec} <\zlim$; (3) $|z_{\rm pri} - z_{\rm sec}| < 500 \,{\rm \kms}$.  We replicate
this procedure 250 times in order to account for variations due to observer
positions.  In later sections, we derive simulation quantities for each of these realizations separately, and use the median and 68 \% limits from this ensemble.

In order to test the fraction of projected satellites that are true satellites
-- i.e., are physically associated with their host, and are not simply close in
projection -- we also compute the true distance to each identified
satellite, and the true distance to the satellite that is closest in 3D (not
in projection).  These are often, but not always, the same (see the next sections).

\subsection{Isolation Criterion and Completeness}
\label{sec:iso}

In comparing our simulated galaxies to our SDSS sample we make use of a mock galaxy catalog that we select to experience the same selection effects as the real data.  In this sense, we do not need to worry about issues like completeness if our only aim is to test this particular \LCDM{} model.  Nevertheless, we can use the simulations to estimate to what extent the selection criterion we impose actually select the galaxy halos we are attempting to select, and to determine the degree of completeness we obtain.

For the SDSS sample, as described in \S \ref{sec:obs}, we require a means of selecting primary galaxies to ensure they are isolated galaxies rather than in a group or galaxy cluster.  To this end, we consider an isolation criterion motivated by \citet{barton07} such that the primary has no other primary within the radius in which we search for secondaries, and at most one out to a larger radius (criterion \ref{crit:iso} from \S \ref{sec:obs}).  This criterion is intended to eliminate cluster galaxies and interacting primaries, while keeping galaxies that are isolated or in very loose groups like the LG. 

Note that any isolation selection will necessitate a trade off between purity and completeness \citep[e.g.][Figure 1]{liu10}.  Our primary concern is purity, and as illustrated below, our choice of sample selection does indeed perform well in that regard \citep{berrier06,barton07,berrier11}.  Purity is the goal because we aim to measure the \emph{fraction} of primaries with a secondary.  Therefore, the overall completeness is not important for the comparison of observations to simulations if our incomplete selection is unbiased.  Our sample does present one possible bias, however: because \LCDM{} features ``assembly bias'' \citep{gao05,wech06,croton07}, sampling from isolated halos in low density regions may result in satellite distributions offset from a more general sample.  Fortunately, as we show later (\S \ref{sec:velcomp}), any assembly bias that may be present because of our selection
 does not seem to affect the radial profiles of dark substructure in the simulations in a significant way.

\begin{figure}[t!]
\epsscale{1.17}
\plotone{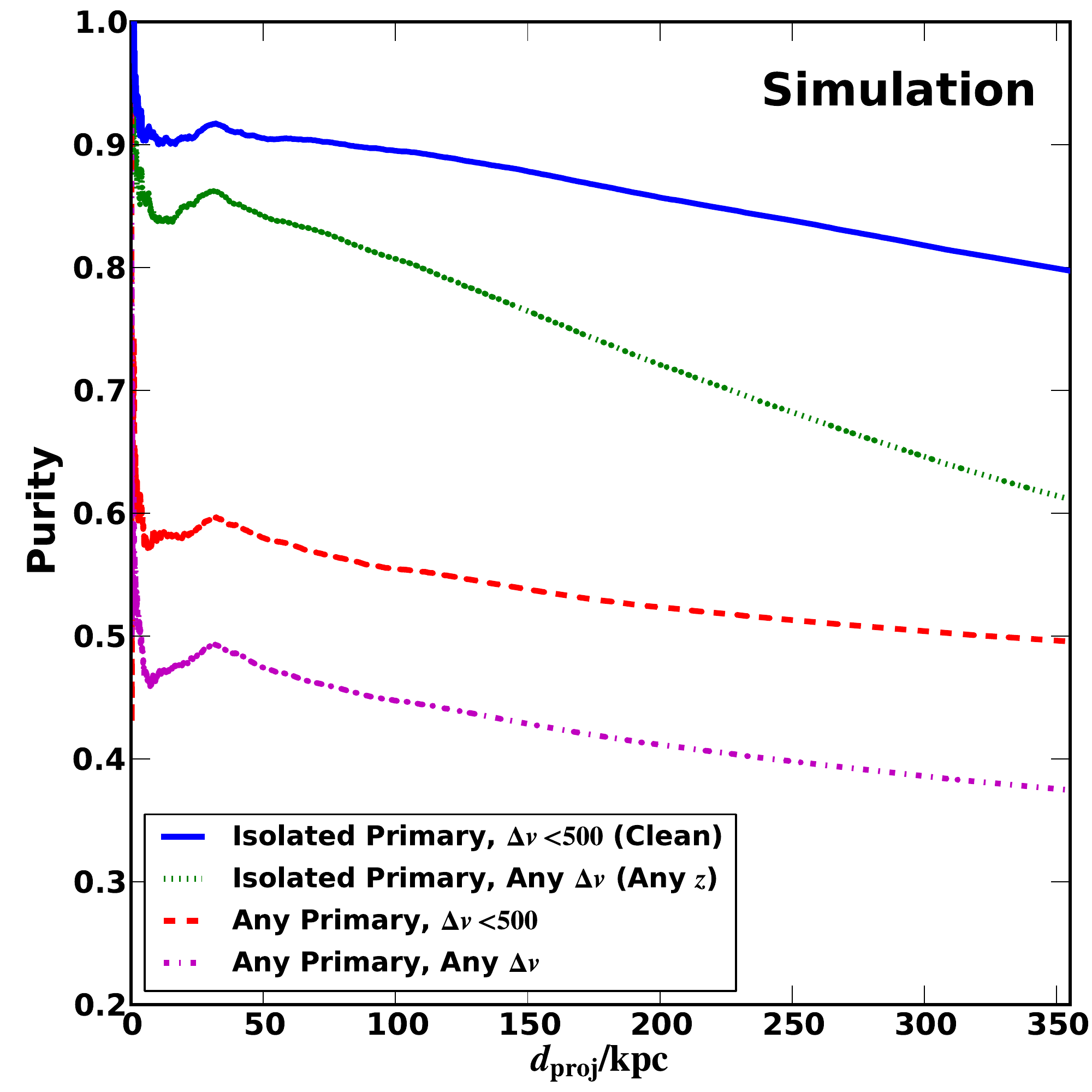}
\caption{Purity of secondary sample in \MSII{}.  Each line shows the fraction of secondary that are actual satellites of the primary (subhalos in a FOF sense) as a function of projected separation.   The solid (blue) line is for the simulated clean sample, green (dotted) is for the any $z$ sample (no $\Delta v$ restriction).  Dashed (red) is for a sample like the clean sample but without the isolation criteria applied to the primary, and dash-dotted (magenta) is any $z$ without isolation.  }
\label{fig:fractrue}
\end{figure}
 
 With a \MSII{} sample (the Simulated Clean sample) selected to match the SDSS sample, we can examine how effectively these isolation criteria function to provide a sample of real satellites. 
Figure \ref{fig:fractrue} shows the {\em purity}: the fraction of primaries in \MSII{} for which the secondary is an actual satellite (a real subhalo of its host)
 as a function of projected primary-secondary distance.  The different lines illustrate how the purity changes for different sample selections.  The solid (blue) line  shows that our fiducial ``clean" selection (isolation criteria plus $\Delta v$ condition) provides a sample that is quite pure, with $80-90\%$ of identified secondaries being actual satellites.  If the isolation criteria are dropped (dashed red) but the $\Delta v$ condition is kept, the purity plummets to  $50-60\%$, suggesting that by selecting isolated primaries we are greatly enhancing our ability to interpret the data.   The dotted green line shows that the $\Delta v$ cut is most important at large separations -- even when we select primaries that are isolated by our criterion, without a $\Delta v$ condition the purity drops to $\sim 65 \%$ at the $250$ kpc projected.  As illustrated by the dot-dashed (magenta) line, less than half of the sample is pure in the absence of either criterion.   In summary, our fiducial clean criteria provide a sample that is made up of $\sim 85\%$ true satellites within $250$ kpc.  The remaining 15\% are halos that are in the secondary mass range, but are isolated (field galaxies) and not actually subhalos of the primary\footnote{The quantitative results here are only valid for the choice of halo and subhalo definitions used by \MSII{} (See \S \ref{sec:sims}).  Other halo-finding methods may result in quantitatively slightly different halo properties, but are unlikely to result in different qualitative results \citep{knebe11MAD},  particularly given how the subhalos we consider here are isolated and  well-resolved.}.

A final question is the incompleteness due to the effect of projection in determining the {\em closest} bright satellite.   Some fraction of the nearest neighbor galaxies in projection will not be the true closest satellites in 3D distance.  While this is not important for comparing \LCDM{} to observations, this does affect the observationally implied fraction of primaries with a true satellite.  We can use the \MSII{} mock survey to estimate this effect in our simulated clean sample by simply computing the fraction of primaries with a secondary within a given projected distance that have a true satellite (a subhalo of the appropriate mass) within the the primary's halo.  Dividing the clean-sample purity (Figure \ref{fig:fractrue}) within the same projected separation by this completeness thus gives the actual fraction of true closest satellites implied by an observed fraction within that separation.  The associated ratio is larger than 2 for any separation within 100 kpc, implying that the projected nearest neighbor is unlikely (less than $1/2$ of the time) to be the true nearest neighbor at separations smaller than this.  This is because the smallest line-of-sight separations tend to be dominated by chance orientation.   Interestingly, the correction factor is unity at a projected distance of $\sim 250$ kpc, implying that the observed fraction of nearest projected galaxies (in the clean sample) should be equal to the true fraction of nearest satellite galaxies (within the three-dimensional virial extent of the halo) at this distance.

\section{Comparing the Observations to the Simulations}
\label{sec:sampanl}

Having established the particulars of our observational sample and identically selected comparisons in the \MSII{} simulation, we can consider whether the observed satellites are consistent with this (bare bones) \LCDM{} model. We do this by comparing the observed galaxy and simulation halo samples through unambiguous measurements in both the observations and simulations:  sky coordinates and line-of-sight velocities.  We also compare luminosity functions for our observed sample to those implied by abundance matching in the simulation.  Additionally, we show this sample does not deviate strongly from the general halo population for $\sim L_*$ galaxies, and infer the 3D satellite distribution.

\subsection{Radial Profiles}
\label{sec:radcomp}

An important comparison between the observations and simulations is that of the projected spatial separation between the primaries and their secondaries \citep[e.g.,][]{chen06,KP09,liu10}.
In the figures that follow, we plot the \emph{fraction} of isolated primaries that have a secondary within a given projected distance $f_{\rm sat}(d_{\rm proj}) \equiv N_{\rm sec}(<d_{\rm proj})/N_{\rm pri}$.  Using the fraction rather than the raw number is preferred as it will be less sensitive to the effects of cosmic variance and the effects of cosmology on the overall abundance of dark matter halos.

\begin{figure}[th!]
\epsscale{1.17}
\plotone{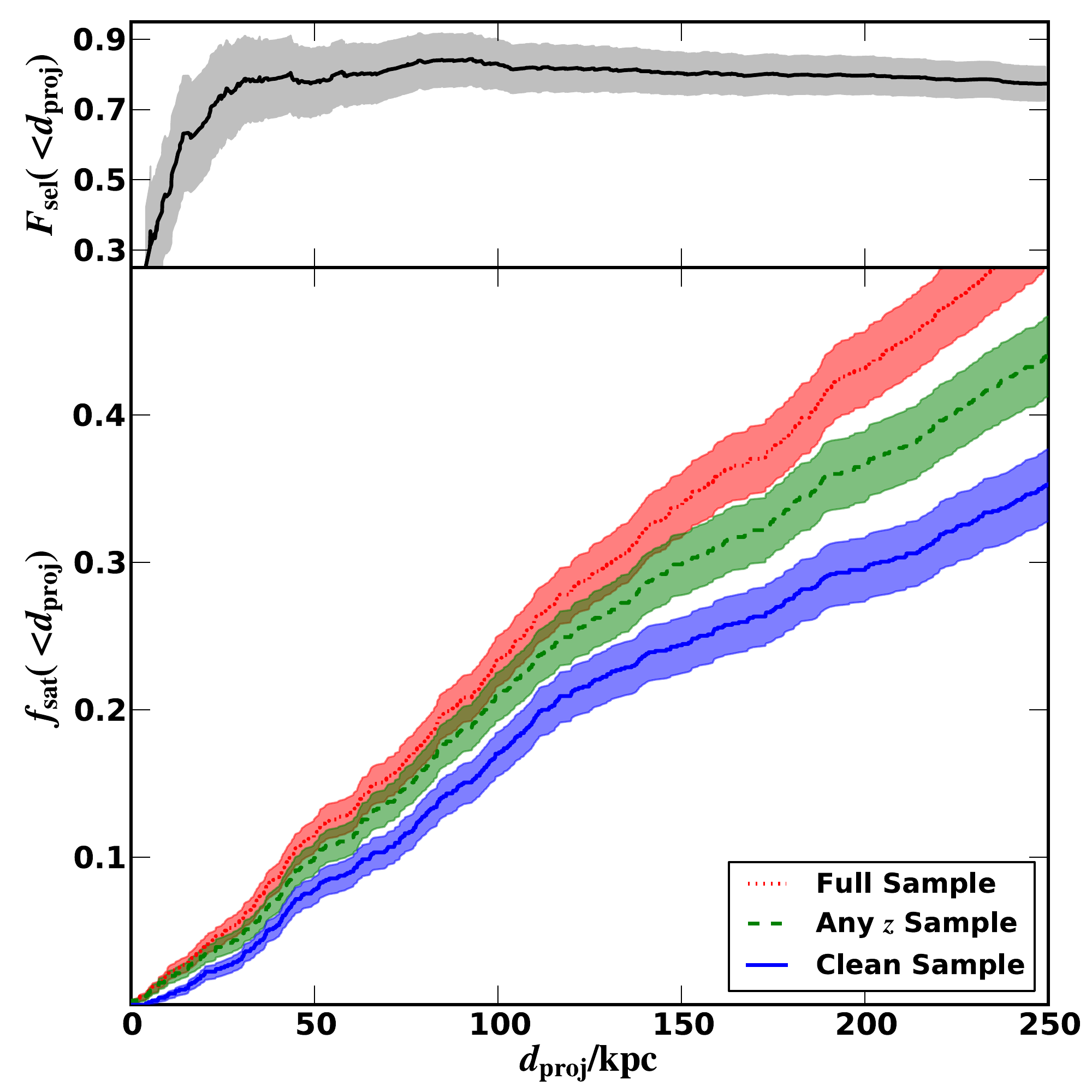}
\caption{Lower panel: Fraction of isolated primaries with a secondary within a particular projected distance for the uncorrected SDSS samples.  The shaded regions are the uncertainties assuming Poisson statistics.   The solid (blue) line is for the clean sample, the dashed (green) is for the any $z$, and dotted (red) is for the full sample.  See Table \ref{tab:samps} for a description of these samples.  Upper panel: Selection function for spectroscopic sample, $F_{\rm sel} = f_{\rm clean}/f_{{\rm Any} z}\rm$ with associated Poisson errors.  This selection function is applied to the full sample for following plots.}
\label{fig:fracradsamps}
\end{figure}

In the lower panel of Figure \ref{fig:fracradsamps}, we plot this fraction for our observational samples.  Solid (blue), dashed (green), and dotted (red) lines show the distributions for the Clean, Any $z$, and Full sample, respectively. See Table \ref{tab:samps} for a description of these samples.  We also show shaded regions around these lines indicating error bars assuming Poisson statistics.

Because the SDSS spectroscopic survey has missed some secondaries due to failed redshifts or fiber collision, the SDSS spectroscopic survey suffers from incompleteness that is disjoint from the biases we explored with the \MSII{} simulation in the previous section.   To correct for this incompleteness we make a simple assumption in the SDSS data: that the true selection function relative to the full sample is the same as the the clean sample relative to the any $z$ sample. We show this selection function $F_{\rm sel} \equiv f_{\rm clean}/f_{{\rm Any} z}\rm$ in the upper panel of Figure \ref{fig:fracradsamps} with the associated Poisson errors; it allows us to determine the true clean fraction in the SDSS sample by taking $f_{\rm sat} \approx f_{\rm clean,corr} \equiv F_{\rm sel} f_{\rm full}$.

\begin{figure*}[htbp!]
\epsscale{1.17}
\plottwo{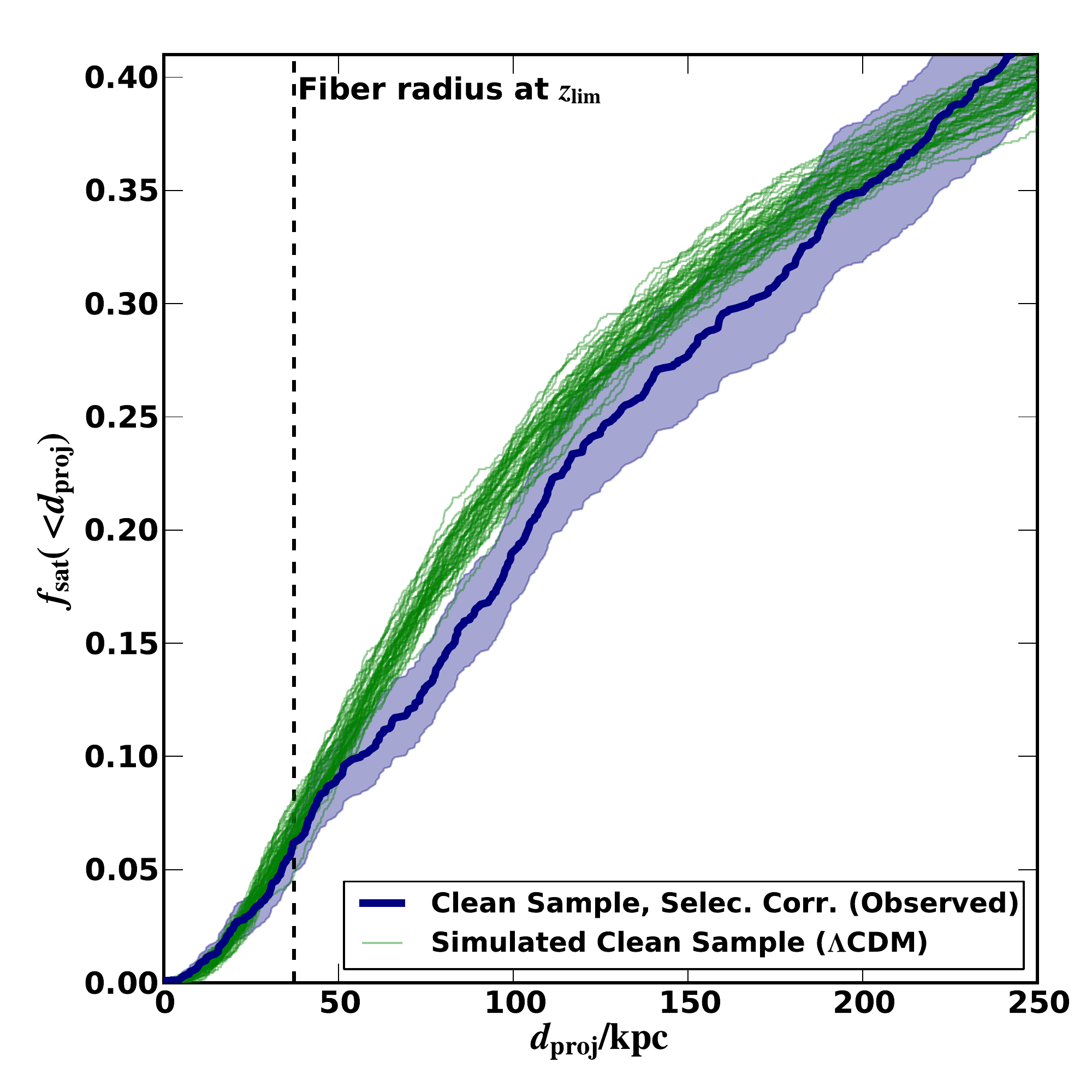}{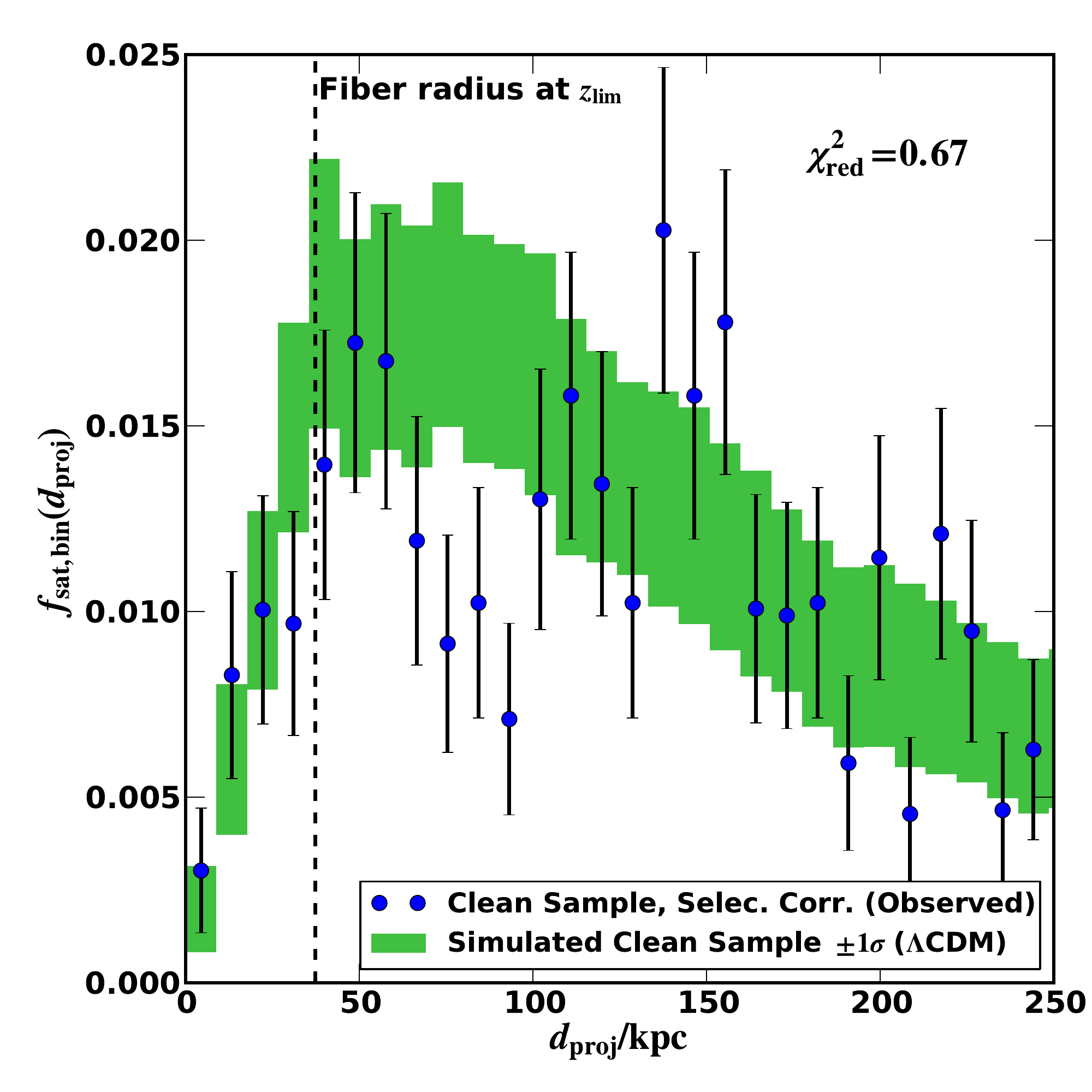}
\caption{Fraction of isolated primaries with a secondary within a particular projected distance in cumulative (left panel) and differential (right panel) form.  The thick blue line on the left panel is for SDSS galaxies in the clean sample, corrected for spectroscopic selection ($f_{\rm clean,corr}$, see text), with Poisson error bars indicated by the shading above and below the line.   The thin green lines are the cumulative fraction of the simulated clean sample for 250 lines of sight in \MSII{} (see \S \ref{sec:mockgal}).  This operates as an estimate of the effect of cosmic variance. The vertical dashed (black) line indicates the distance corresponding to an SDSS fiber diameter at the maximum redshift of the sample, and also the approximate distance at which baryonic effects may render the \MSII{} sample uncertain.  The right panel presents the same information in differential/binned form, giving the fraction per bin in $d_{\rm proj}$.  Points (blue) with error bars are for the SDSS clean sample and with Poisson errors, and the (green) bars are $1\sigma$ ranges for the \MSII{} sample lines of sight.  The reduced chi-squared value comparing the observations to simulation is also shown in this panel, and it indicates no statistically significant deviation between the two (if the bins are uncorrelated). For both panels, the observed fraction is corrected for selection as described in \S \ref{sec:sampanl} by multiplying the fraction in the full sample by $F_{\rm sel}$ of Figure \ref{fig:fracradsamps}.}
\label{fig:fracrad}
\end{figure*}

The corrected nearest secondary fraction, $f_{\rm clean,corr}$, is plotted as a function of projected separation in the left panel of Figure \ref{fig:fracrad}  as the solid (blue) line with errors due to Poisson statistics shown as the blue shaded region.  Over this, we plot the \MSII{} Simulated Clean sample for all 250 lines of sight as the thin green lines.  We also show the fiber radius at the maximum redshift of our sample as the vertical dashed (black) line.  This radius is also the approximate \emph{physical} (rather than projected) separation at which baryonic effects not included in \MSII{} render the simulation less reliable, so we consider results inside this distance subject to additional uncertainty that is difficult to quantify.    The same data are plotted in differential form in the right panel of Figure \ref{fig:fracrad}.  The number of SDSS galaxies in each $d_{\rm proj}$ bin is given by (blue) points with Poisson error bars, and the \MSII{} galaxies are given by (green) bands representing $1\sigma$ scatter over all lines of sight.  From this plot, it is clear that the SDSS clean sample and \MSII{} match over most of the range, although with a moderately significant dip at $\sim 80$ kpc.  Further, we compute the chi-squared for these data, and find $\chi^2_{\rm red}=0.67$, with a probability $p= 94 \%$ of a $\chi^2$ this high or higher if the simulation is a model for the observations.  These results indicate a remarkable consistency between the SDSS observations and the predictions of \LCDM{}.  There is no overabundance (nor underabundance) of luminous satellites of isolated $L_*$ galaxies.

Before moving on, we draw attention to the fact that we have only discussed the projected separation distribution for $d_{\rm proj} < 250$ kpc.  This is intentional.  
As mentioned in \S \ref{sec:obs}, the targeting scheme of the SDSS spectroscopic survey introduces a bias at radii larger than this that is difficult to quantify accurately.  If a background cluster (or group) is close enough to a primary that it is within the projected distance expected for a secondary, the cluster galaxies will be considered as possible secondaries.  Because there are multiple bright neighbors in the cluster, only one will be assigned a fiber by the SDSS tiling algorithm, and the others will have no redshift.  If assigned the redshift of the primary, often at least one of these galaxies will fall into the magnitude range of a secondary.  Hence, in the full sample (Table \ref{tab:samps}), they will be designated the secondary for that primary if there is no other nearer secondary.  Thus, the background galaxy is taken to be a secondary, excluding a true satellite that could be present at a larger projected distance. This effect is not fully accounted for by the use of $f_{\rm clean,corr}$ because  the number of potential secondaries in the no $z$ sample increases by the number of cluster galaxies without redshifts, while the any $z$ sample only increases by one.   At large separations, this effect becomes significant: at  $d_{\rm proj} \gtrsim 350$ kpc, many secondaries in the no $z$ sample are visually associated with background cluster galaxies. Fortunately, the presence of a background cluster depends strongly on $d_{\rm proj}$ due to the $d^2$ scaling of background volume. Visual inspection of candidate secondaries at large separation in the full sample shows that this effect does not become significant until  $d_{\rm proj} \gtrsim 250$ kpc.
Another reason we have chosen to limit our analysis to $d_{\rm proj} < 250$ kpc has to do with the analysis of purity and completeness described in \S \ref{sec:iso}: the completeness and purity corrections conspire to be $\sim 1$ at that radius, so the fraction at $250$ kpc in Figure \ref{fig:fracrad} can be interpreted as an estimate of the true fraction of satellites within the virial extent of the halo.  Thus, we expect that $\sim 40 \%$ of isolated $\sim L_*$ galaxies have a luminous (LMC-like) satellite within their dark matter halo \footnote{The exact fraction is a function of the host halo mass, as discussed in \S \ref{sec:halobias} and \S \ref{sec:comparison}.}.

\subsection{Pairwise Velocity Distributions}
\label{sec:velcomp}

\begin{figure}[t!]
\epsscale{1.27}
\plotone{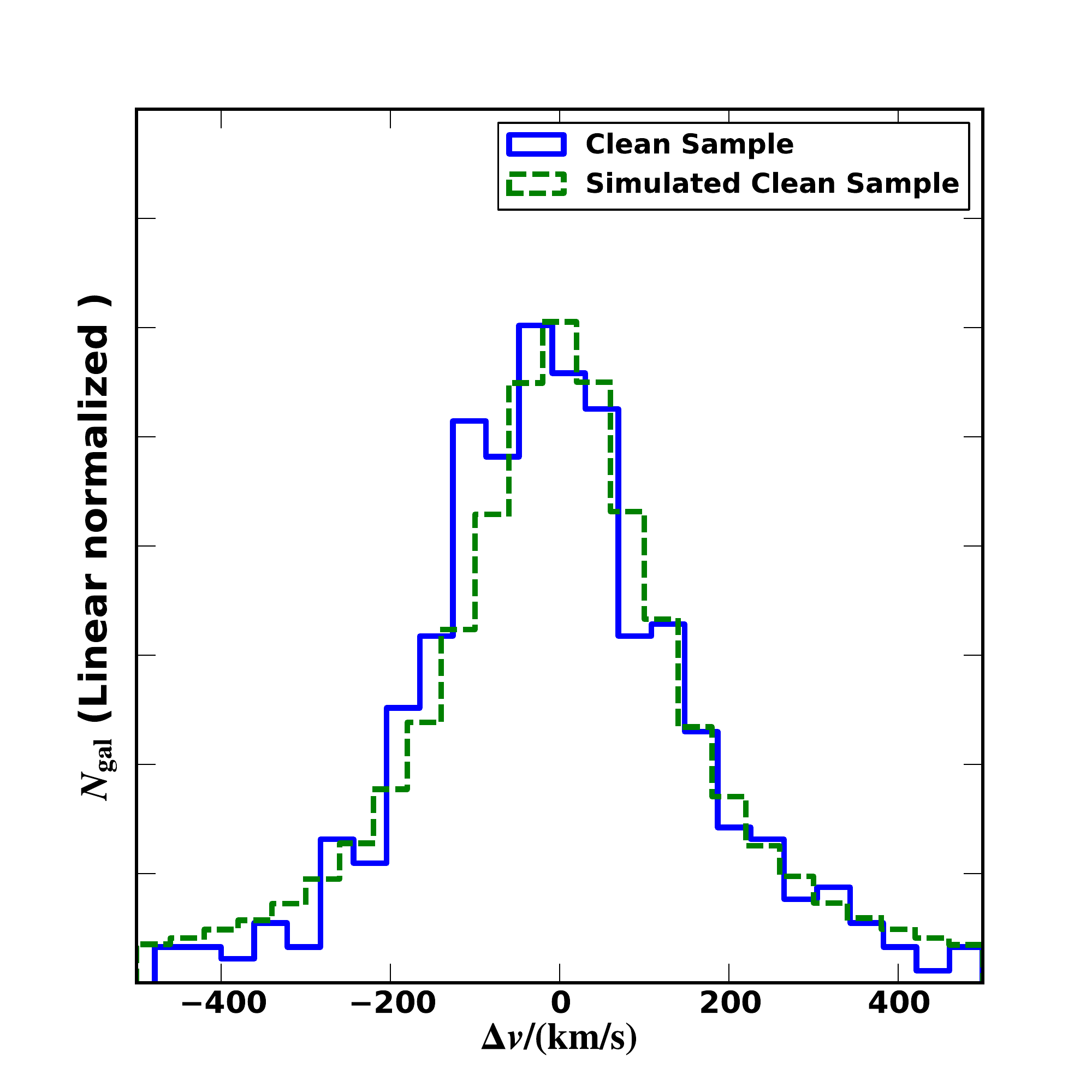}
\caption{Distribution of $\Delta v \equiv c(z_{\rm pri}-z_{\rm sec})$ for the clean sample (solid blue histogram), the clean-like sample from \MSII{} (dashed green).  The KS test yields $\pks{33}$. The pairwise velocity dispersion in the observed sample is $\sigma=161$ $\kms$. }
\label{fig:dvdists}
\end{figure}

The line-of-sight velocities for the SDSS galaxies are directly determined from the measured redshifts of the galaxies in the SDSS sample, and they are easily measured for \MSII{} halos by projecting the halo velocity along the line-of-sight.  In Figure \ref{fig:dvdists}, we show the pairwise velocity distribution $\Delta v \equiv c(z_{\rm pri}-z_{\rm sec})$ for the clean sample and the simulated clean sample.  We compute the dispersion for the observed sample and find $\sigma=161$ $\kms$, of interest as a possible tracer of host or cosmological properties \citep[e.g.][]{strauss98}.  The pairwise velocity dispersion also provides an indirect test of the halo mass-luminosity assignment that we have used in our simulation (e.g. if our halos were too massive then our simulated velocity distribution would presumably be too large).  
As illustrated in 
Figure \ref{fig:dvdists}, the simulated and observed distributions agree quite well.  We also compute the Kolmogorov-Smirnov (KS) statistic between the two distributions.  This test shows that the simulation and observations are consistent, with a $\pks{33}$ probability of the observed KS D statistic if they are drawn from the same distribution.  This result is consistent with no velocity bias between the galaxies and their subhalos \citep[e.g.][]{blanton99bias}, although it says nothing about bias of the subhalos relative to the smooth host dark matter distribution \citep[e.g][]{colin00bias}.  Thus, the relative line-of-sight velocities in SDSS for likely bright satellites around $L_*$ hosts are fully consistent with the predictions of \LCDM{}.

\subsection{Luminosity Function}

\begin{figure}[t!]
\epsscale{1.18}
\plotone{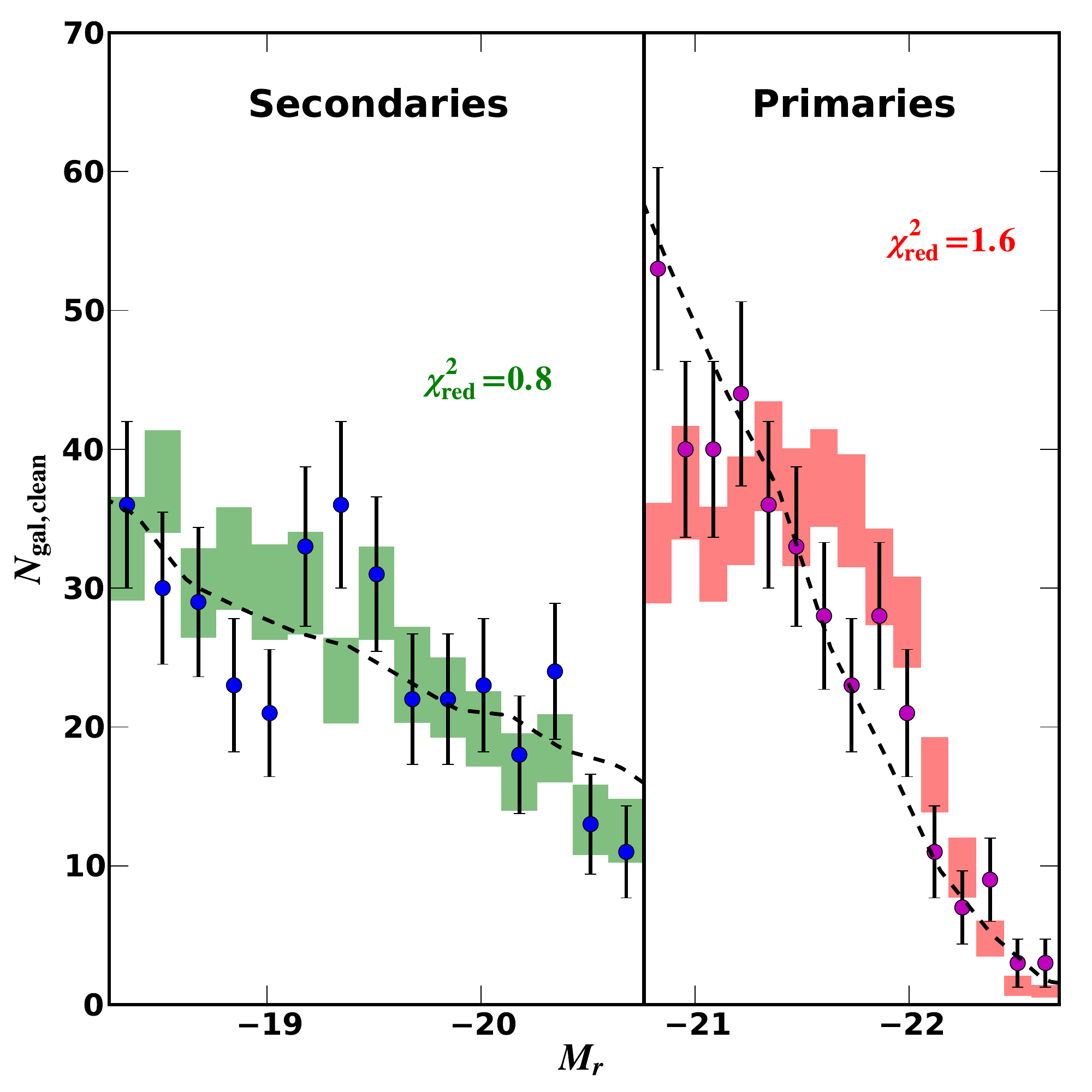}
\caption{Comparison of luminosity functions.  Brighter than $M_r<\prilim$, the luminosity function for clean sample primaries is shown as (magenta) points, and selection corrected clean sample secondaries are shown on the faint end as (blue) points.  Error bars on the points are Poisson.  Shaded bars are for the 68 \% bounds of the 250 realizations of the simulated clean sample, again for primaries on the bright end (red) and secondaries fainter than $M_r<\prilim$ (green).  Luminosities have been assigned to subhalos with abundance matching, as described in \S \ref{sec:abundmatch}. Also shown are reduced $\chi^2$ values for a $\chi^2$ test comparing the clean and simulated clean samples.  Dashed black curves are the luminosity function from \citet{blanton05dwarflf}.  The simulated clean and \citet{blanton05dwarflf} curves are scaled to match the normalization of the clean sample. 
}
\label{fig:lfuncs}
\end{figure}

In \S \ref{sec:radcomp}, we show that the \MSII{} mock galaxy catalogs (and by proxy, \LCDM{} predictions) match the observed projected separation distribution of luminous satellites.   Here, we compare the luminosity function implied by the observed and simulated samples.  While a luminosity function of our volume-limited observational sample is straightforward to compute, comparison to the simulation is complicated by the need to use abundance matching to map galaxy luminosities to halos.  We proceed as described in \S \ref{sec:abundmatch}, but assign a luminosity to each halo instead of just computing mass ranges for the primary and secondary samples.  The resulting luminosity functions are shown in Figure \ref{fig:lfuncs} for the selection corrected clean sample (points) and the simulated clean sample (bars).  

We do not directly compare the normalization of the luminosity functions, as the selection of primary galaxies for the SDSS spectroscopic survey results in a highly uncertain survey volume (motivating use of the primary \emph{fraction} in \S \ref{sec:radcomp}).  Instead, we adjust the normalization of the simulated clean sample to match the observed clean sample and compare shapes. As the reduced $\chi^2$ values in Figure \ref{fig:lfuncs} indicate, the luminosity function for secondaries matches well within the error bars, while the primaries are marginally deviant.  This level of agreement is noteworthy, as subhalo abundance matching is not well constrained at these low masses/luminosities.  While it is highly effective where calibrated for the full galaxy sample \citep[][and references therein]{krav04hod,CW09,moster10}, abundance matching  may not be as effective for biased subsets of galaxies \citep[e.g.][]{toll11}.  Thus, the disagreement for the primaries may be due to a preference for a particular subclass of more or less massive galaxies (e.g. red/passive vs. blue/starforming) to host secondaries.  The abundance matching does not capture the differing luminosity functions of these two classes and hence would not match the observed sample. Alternatively, the bias described at the end of \S \ref{sec:abundmatch} may account for the disagreement, as it would manifest as an enhancement in the observed sample in the faintest bin.  Removing this faintest bin lowers the reduced $\chi^2$ to 1.1, suggesting it may be the primary effect.

The luminosity functions in the simulation and observations agree reasonably well, further suggesting that \LCDM{} matches the observations.  However, this is not necessarily a strong test compared to that discussed in the previous subsection.  The (black) dashed lines in Figure \ref{fig:lfuncs} show the luminosity function of \citet{blanton05dwarflf} in the primary and secondary regimes.  It is clear that the shape of these luminosity functions roughly match the clean observed and simulated samples.  Thus, this primarily shows that the clean samples very roughly trace the overall luminosity function (the \emph{overall} luminosity function implied by the simulations must match by construction).  This is, however, a necessary consistency check that these samples (and hence \LCDM{}) pass.

\subsection{Host Halo Sample Biases}
\label{sec:halobias}

We have shown that the isolated (clean) SDSS sample matches the simulated clean sample from \MSII{} remarkably well.  We now ask whether our isolation criteria have forced us to select a special set of host halos that are not representative of the average dark matter halo in their satellite properties.  That is, have we measured the satellite statistics that are representative of all $\sim L_*$ halos or just a special group of isolated $\sim L_*$ halos?  Such a concern is  motivated by the possibility of assembly bias \citep{gao05,wech06,croton07}.

\begin{figure}[t!]
\epsscale{1.17}
\plotone{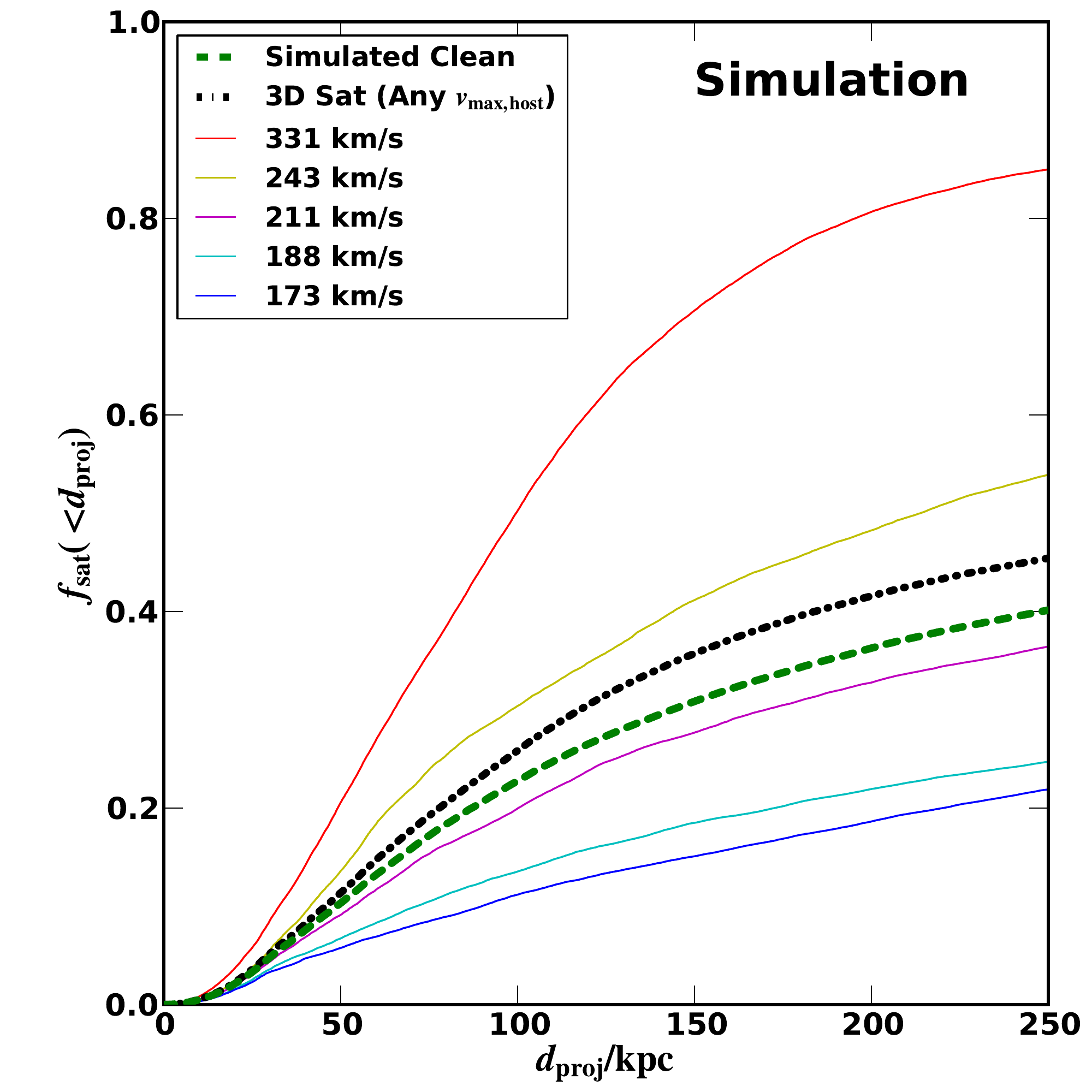}
\caption{Fraction of simulated primaries with a secondary within a given projected distance compared to a more general set of host halos.  The dashed (green) line is for the simulated clean sample and its secondaries.  The dot-dashed (black) line is the distribution of the \emph{3D nearest} object (nearest in $r$ rather than $d_{\rm proj}$) for all hosts in the \MSII{} SDSS-like volume.  This sample is not subject to assembly bias effects in the way the clean sample may be, as it is not explicitly selected to be isolated.  The thin solid lines show fractions for the same hosts sub-divided into quintiles in $\vmax$ ($v_{\rm max,host}$ decreasing from upper-most to bottom line).  The median $v_{\rm max,host}$ values for each line are indicated in the figure legend. }
\label{fig:fracisocosmo}
\end{figure}

In Figure \ref{fig:fracisocosmo} we use the \MSII{} simulation to plot the fraction of primaries with a secondary within a particular projected separation for a more general sample of halos.  The Simulated Clean sample (also shown in Figure \ref{fig:fracrad}) is plotted as the dashed (green) line.  The dash-dotted (black) line plots the separation distribution for \emph{all} host halos in the \MSII{} SDSS-like volume in the primary mass range, regardless of whether or not they meet our isolation criteria.  Additionally, secondaries are assigned by identifying the nearest object in 3D (not projected) distance, to eliminate the projection effects described at the end of \S \ref{sec:iso}.  We also split this sample into quintiles by host halo $\vmax$ and plot these distributions as the thin solid lines in Figure \ref{fig:fracisocosmo}.  As described in more detail in \S \ref{sec:comparison}, examining these lines clearly shows that the exact fraction at a given separation is strongly dependent on the choice of mass ranges used for the primary, and hence care must be taken comparing similar samples in other works to ensure they are closely-matched in primary/host properties.

While there is a discrepancy of $\sim 4\%$ at 250 kpc between the Any $v_{\rm max,host}$ distribution (black dash-dotted) and the Simulated Clean (green dashed), this is only marginally larger than the scatter over multiple pointings in the mock galaxy catalogs, and is $<1\sigma$ compared to the Poisson errors in the \emph{observational} sample.  Additionally, the mass bin centered around the median $v_{\rm max,host}$ for the abundance-matched sample (211 km/s; magenta line) is even closer to the Simulated Clean distribution.  Hence, while the SDSS clean sample at first appears rather tuned and specific in its parameters, it is in fact a fairly general sample that is representative of true satellites of $\sim L_*$ hosts.  Specifically, there is no sign that differences in halo assembly time associated with large-scale clustering have given rise to any observable bias in our isolated clean sample compared to the general halo population.

\subsection{3D Density Distribution of Satellites}

We now turn to an estimate of the 3D distribution of satellites directly from the data themselves.  We start with the 2D (selection-corrected, $f_{\rm clean,corr}$) clean profile from the right panel of Figure \ref{fig:fracrad}, and determine the on-sky surface density in these annuli.  From this surface density, we numerically perform an inverse Abel Transform to approximate the 3D density profile under the assumption of spherical symmetry\footnote{While the satellite distribution of any individual host is not necessarily spherically symmetric, the halos are oriented randomly relative to the observer in an isotropic universe, and thus the ensemble average for all primaries is spherically symmetric. }.

\begin{figure}[t!]
\epsscale{1.17}
\plotone{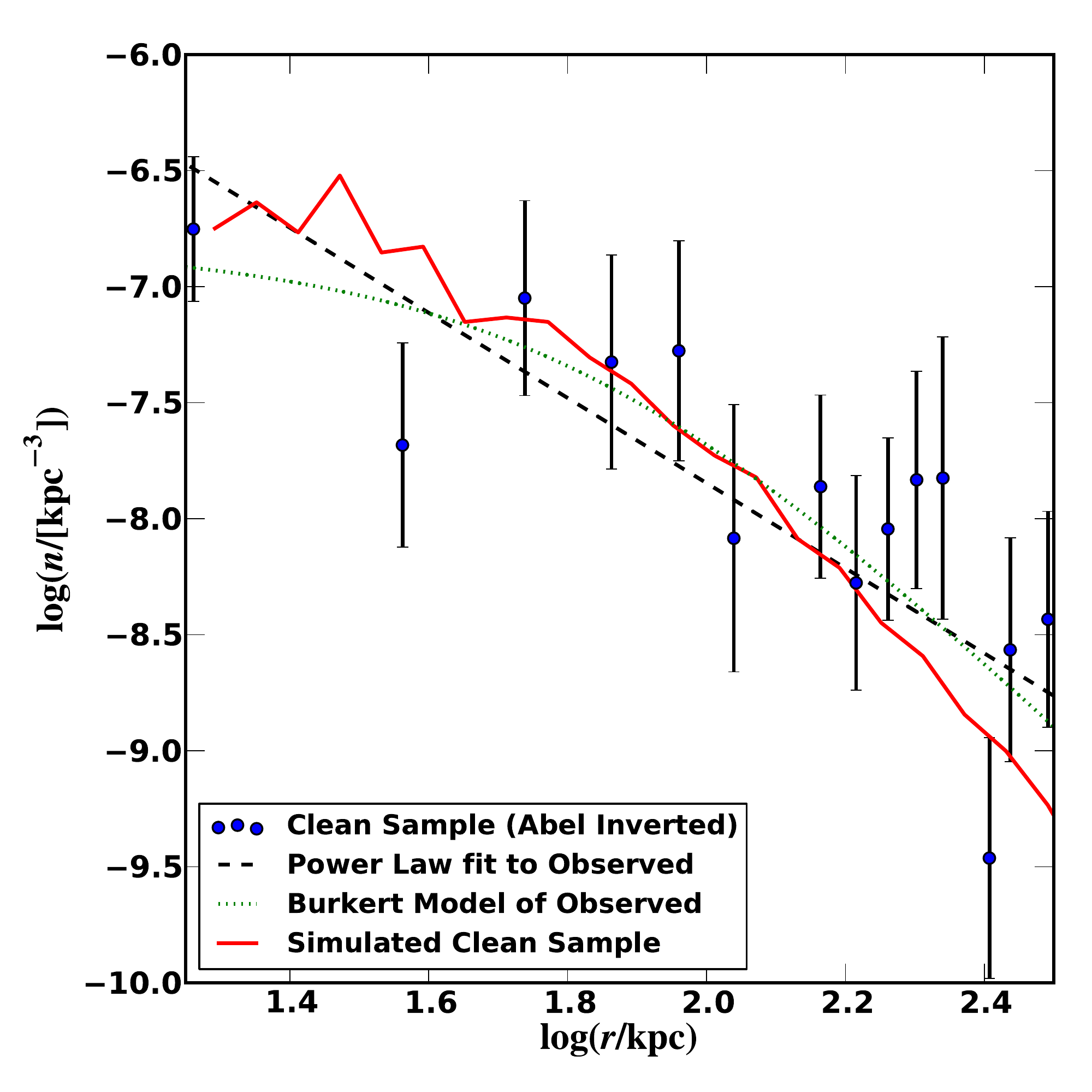}
\caption{Number density profile (3D) of nearest satellite.  Points (blue) are computed as the Abel Transform of the observed selection corrected clean sample separation distribution, normalized to give a fraction of 42\% at 250 kpc.  The error bars are generated by a combination of Poisson errors and bootstrap resampling (with replacement).  The dashed (black) line is the best fit power law ($\rho \propto r^{-1.8}$) for this data set, normalized to give the same fraction integrated over the range shown.  The dotted (green) line is a \citet{burkert95} profile fit to the observed data, with core radius $r_0=68$ kpc and $n_0 = 1.6 \times 10^{-7} {\rm kpc}^{-3}$.  Finally, the solid (red) curve is the 3D density profile of the simulated clean sample from \MSII{}.
}
\label{fig:threedprof}
\end{figure}

The resulting 3D number density profile obtained by this prescription is shown in Figure \ref{fig:threedprof} as the (blue) points.  This profile is by necessity coarsely binned to reduce numerical noise due to small number statistics, and the missing bins at small radii are due to the declining number of secondaries at radii inside the fiber radius.  We also show error bars obtained from a combination of Poisson error and bootstrap resampling to simulate other choices of binning.  We fit a power law to the data (black dotted line), normalized to integrate to 42\% from the inner most point to 250 kpc, obtaining a slope of -1.8 (and normalization $6.6 \times 10^{-5}$).  Also shown is the 3D density profile of satellites of the simulated clean sample (green solid line), normalized to match.  The density profile in the simulations transitions from a nearly flat core to a profile falling off as roughly $r^{-3}$.  This profile is consistent with the Abel inverted observed profile, motivating a fit to a \citet{burkert95} profile as an approximation to the observed profile.  The resulting fit has a core radius $r_0=68$ kpc, or $\log(r_0/{\rm kpc})=1.8$, and central number density $n_0 = 1.6 \times 10^{-7} {\rm kpc}^{-3}$.  These estimates of the profile, however, come with the crucial caveat that the shape (particularly the presence of a core) cannot be truly constrained by the observations without a much larger data set that lacks the systematic biases induced by fiber collision.

\section{Photometric Distribution of Secondaries and Comparison to the LMC}
\label{sec:lmc}

Having compared the observational sample to the \LCDM{} predictions in \MSII{}, we now overview the photometric characteristics of the secondaries.   In this section we focus on the clean sample, as it is the sample that has the highest fraction of likely satellites.  This sample is not complete due to instrumental failures and the effect of distant background clusters (see the discussion in \S \ref{sec:sampanl}).  However this incompleteness is primarily random, and hence should not systematically bias the photometric properties of the sample\footnote{While fiber collision may bias the sample away from very close pairs, these are always a small fraction of the sample, as the total volume available for satellites at small separations is very small.}.  

\subsection{Photometric Properties of Secondaries}

\begin{figure}[t!]
\epsscale{1.25}
\plotone{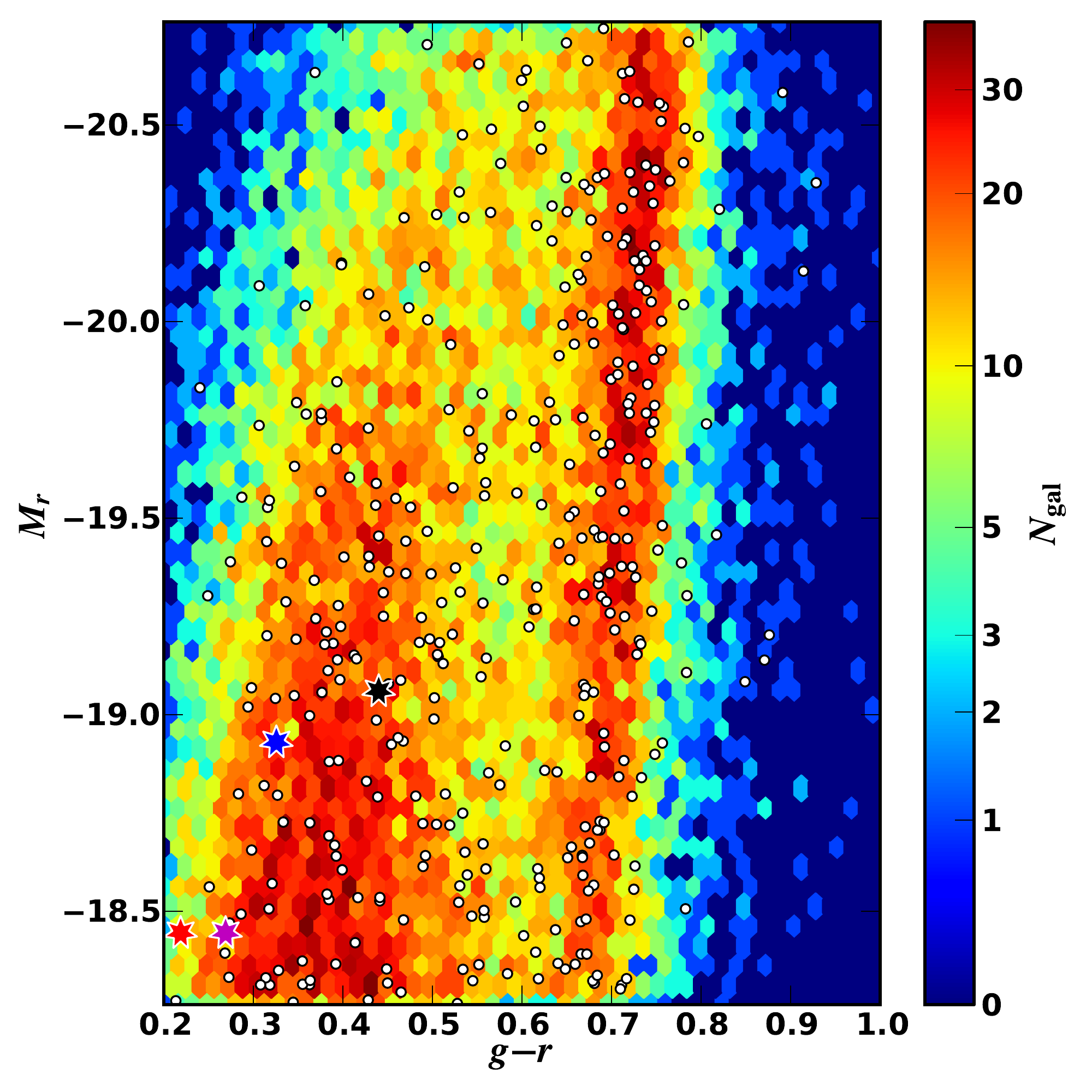}
\caption{Color-magnitude diagram ($g-r$ and $M_r$) for the secondaries with $d_{\rm proj} < 250$ kpc.  White circles are the clean sample,  and the underlying hex-binned plot is a Hess-like diagram indicating the number of SDSS galaxies in each bin (following Figure \ref{fig:matchcmd}).  The LMC is shown as the three star-shaped symbols for the three color estimates: from left to right, RC3 (red), \citealt{eskew11} (magenta), and \citealt{bothun88parkinglot} (blue).  The black-star shaped symbol is for M33 (see text).
}
\label{fig:matchcmdsec}
\end{figure}

\begin{figure}[t!]
\epsscale{1.27}
\plotone{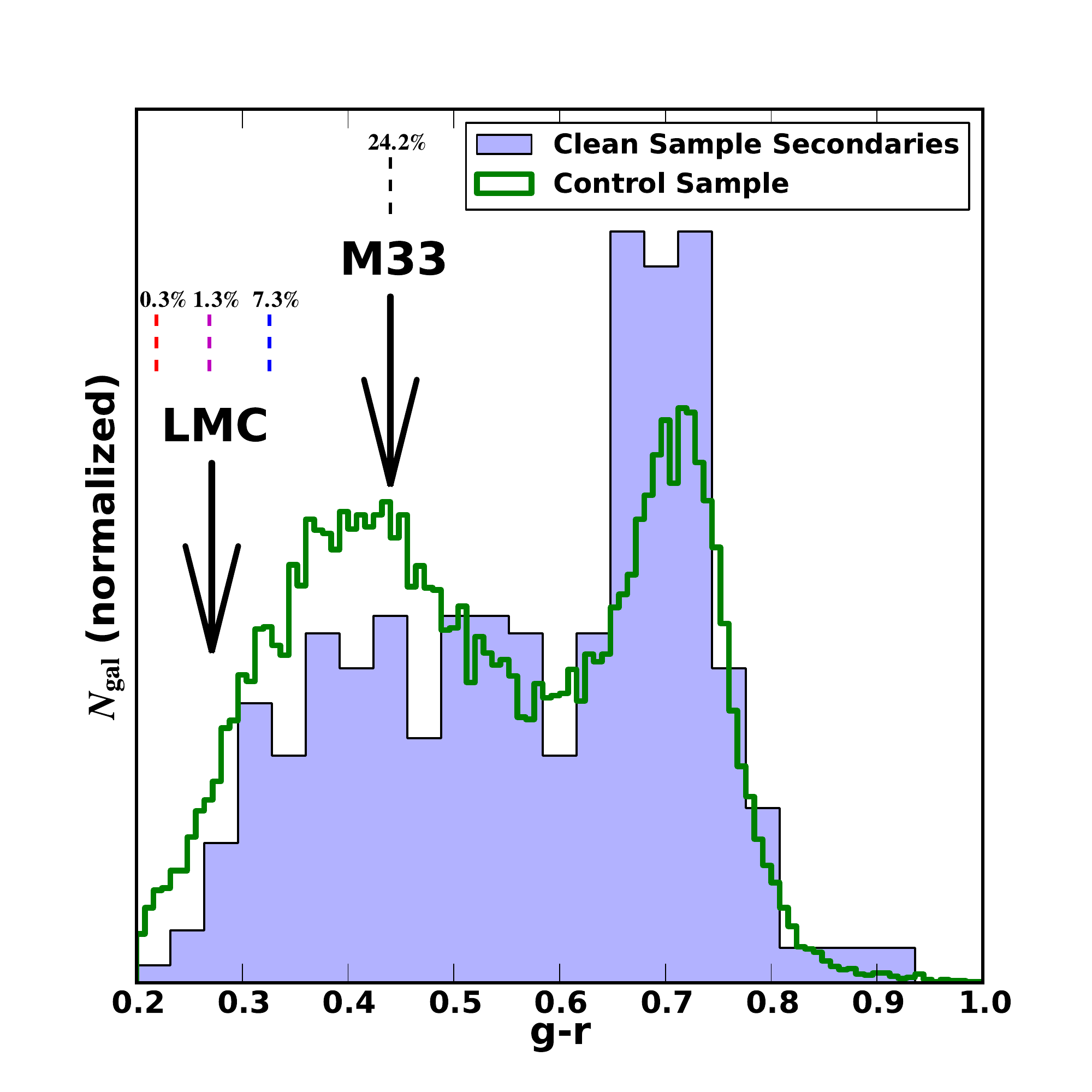}
\caption{Color distribution of secondaries in the observational data set.  The shaded (blue) histogram is for secondaries in the clean sample, while the line (green) histogram is for  the control sample (all objects in the volume limit and magnitude range for secondaries). The KS test yields $p_{\rm KS} \approx 10^{-7}$.  Short dashed vertical lines show LG galaxy color estimates: from left to right, LMC from RC3 (red), \citet[][magenta]{eskew11}, and \citet[][blue]{bothun88parkinglot}; and M33 (black, see text). The $p$ values in the legend give what percentile of the clean sample (the shaded histogram) that color corresponds to. The arrow for the LMC is placed at the mean of the three color measurements.}
\label{fig:colorhist}
\end{figure}

In Figure \ref{fig:matchcmdsec} we show the $g-r$, $M_r$ CMD for the secondaries in the clean sample as the white circles, plotted over all galaxies within the volume limit that have magnitudes in the secondary range.  We also show the color distribution in Figure \ref{fig:colorhist} for the clean sample as the shaded (blue) histogram and the Control sample (a collapsed version of the background of Figure \ref{fig:matchcmdsec}) as the solid lined (green) histogram.  The Control sample, which includes both satellites and non-satellites, clearly shows the ``blue cloud''/``red sequence'' color bimodality \citep{strat01} and hence below, when discussing blue cloud galaxies, we refer to galaxies with  $g-r<0.6$.    

The most striking feature of Figure \ref{fig:colorhist} is the fact that the color distribution of the Clean sample secondaries is visibly different from the Control, confirmed by KS test at the $p_{\rm KS} \approx 10^{-7}$ level.  For the secondaries, 52\% are on the red sequence, while in the Control sample, only 38\% of galaxies are red.  This is even more surprising given that our Control sample has galaxies in all environments, not just isolated (low-density) environments.  A control sample selected to be isolated like the primaries would be significantly \emph{bluer} than average due to the density-morphology or density-star formation rate relations \citep{dressler80,gomez03sfrden}.  Because our Control sample is not selected in this manner, the difference between secondaries and similar isolated field galaxies should be even greater than shown here.   

This result implies that star formation is suppressed even for single satellites of isolated $L_*$ host galaxies.  This suggests that harassment by other luminous satellites \citep{moore96harass} cannot be a major cause, and thus other effects such as strangulation or ram pressure stripping  must be at play \citep[e.g.][and references therein]{vdb08}. Given the nearly complete lack of gas for MW satellites other than the MCs, however \citep{grcevich09}, the mechanism driving morphological transformation for these objects may be different from that in denser environments. A more in-depth analysis of the secondaries in this context is warranted but beyond the scope of the current work.

\subsection{Local Group Photometric Estimates}

To determine what the clean secondary sample can tell us about the MW/LMC system, we must estimate the integrated absolute magnitudes and colors of the MW and LMC as the SDSS would observe them from outside the LG.  Our SDSS data set focuses on the $r$- and $g$-bands, where signal-to-noise is best, so below we estimate the $r$-band absolute magnitudes of the MW and LMC, as well as their $g-r$ color.

Determining the  luminosity of the MW in the optical is notoriously difficult due to the high extinction in the disk, as well as the necessity of precise photometry over a large fraction of the sky.  An often-cited estimate from \citet{vdb00lgbook} is $M_V=-20.9$.  To convert to SDSS $r$ band, we determine the average color of a sample of Sbc galaxies \citep[the type for the MW according to][]{vdb00lgbook} in the SDSS from \citet{nair10sdssmorph}, finding $<g-r>_{\rm Sbc} = 0.60$.  We then use the \citet{jester05sdsstrans}  photometric transforms and convert from Vega to SDSS AB magnitudes, yielding  $M_{r,{\rm MW}}=-21.2$. This is within 0.1 mag of the median of the SDSS sample of the primaries (and also $M_*,r$, by design of the sample).

For the LMC, a somewhat different set of complications are present.  Even the high surface brightness parts of the LMC cover a relatively large area on the sky ($\gtrsim 100$ sq. degrees), and hence accurate measurements of the LMC's integrated magnitudes require photometry with consistent background subtraction over the entire area.  Hence, the overall magnitudes for the LMC are rather uncertain.  The color is easier to measure because it does not require consistent calibration over the entire area.  Uncertainties in the distance to the LMC also propagate to uncertainties in the integrated absolute magnitude, but the uncertainties in the surface photometry are much larger.  Thus, we assume a distance to the LMC of 50 kpc \citep[distance modulus 18.50, e.g.][and references therein]{lmcdist1}.

With these complications in mind, we consider the foreground extinction-corrected apparent magnitude from the Third Reference Catalogue of Bright Galaxies \citep[RC3][]{rc3} for our estimate of the LMC's $r$ band absolute magnitude.  The RC3 gives $U,B,V=0.52,0.57,0.13$, which we adjust to SDSS colors using the transformations of \citet{jest05sdsstrans}, yielding $M_{r,\rm LMC}=-18.44$.  The $g-r$ color implied from the RC3 magnitudes is $(g-r)_{\rm LMC} = 0.22$.  

A second determination of the LMC integrated properties from surface photometry results from the work of \citet{bothun88parkinglot}.  These observations were obtained with a CCD camera mounted to a standard camera lens, providing a wide enough field of view to cover a sizable fraction of the LMC in a single exposure.  With foreground extinction correction for the LMC assuming $E(B-V)=0.075$ \citep{SFD98}, these observations and the above transforms yield $(g-r)_{\rm LMC} = 0.33$. While the field of view was still not large enough to cover the luminous extent of the LMC, an extrapolation of an exponential profile yields $M_{r,\rm LMC}=-18.93$.

We also consider an alternative to surface photometry: stellar population synthesis.  This method provides colors by computing the luminosity-weighted contribution of the present-day stellar populations with an age and metallicity distribution based on an empirical star formation history (SFH) and final metallicity.  This method has the advantage of not requiring the careful calibration necessary for surface photometry, but has the disadvantage of being more model-dependent.  We consider here the results of \citet{eskew11} based on the LMC SFH of \citet{harris09lmcsfh}. From these stellar populations, the color of the LMC at the present day is $(g-r)_{\rm LMC} = 0.27$. Adopting the stellar mass implied from the SFH of \citet{harris09lmcsfh} then yields $M_{r,\rm LMC}=-18.44$.  These results differ from the surface photometry; they do \emph{not} include the effects of reddening and extinction internal to the LMC, as they assume all stars in the synthetic population contribute their full luminosity.  

Based on the above measurements, we adopt $M_{r,\rm LMC}=-18.44$ ($L_{r} = 1.75 \times 10^{9} L_{\odot}$) from the RC3 catalog as our fiducial estimate for the LMC, as it is a direct measurement more akin to SDSS observations (although it is consistent within error bars with the \citet{harris09lmcsfh} result), requiring no extrapolation.  While this luminosity is marginally fainter than the overall median of the secondaries in our SDSS sample ($M_r=\medsec$), it \emph{is} close to the median of the blue cloud galaxies apparent in the background of Figure \ref{fig:matchcmdsec}.  Additionally, there is no compelling reason to favor the $g-r$ colors of any of these results over the others, and thus we consider all of them.  Thus, we summarize the photometric estimates for the LMC in Table \ref{tab:phots}.

\begin{deluxetable}{cccc}
\tablecolumns{4}
\tablecaption{LMC Integrated Photometric Properties}
\tablehead{
  \colhead{Source} &
  \colhead{Method} &
  \colhead{$M_r$} &
  \colhead{$g-r$} 
}
 \startdata
\citet{rc3} & Phot.  Surf. Phot. & -18.44 & 0.22 \\
\citet{bothun88parkinglot} & CCD Surf. Phot. & -18.93 & 0.34 \\
\citet{eskew11} & Population Synthesis & -18.44 & 0.27
 
\enddata
\label{tab:phots}
\end{deluxetable}

\subsection{LMC and M33 in Context}

An additional feature of Figure \ref{fig:matchcmdsec} is the location of the LMC using our three color estimates and the RC3 magnitude as the star-shaped symbols.  We also indicate the LMC color estimates as the left three vertical lines in Figure \ref{fig:colorhist}.  For all colors we consider here, the LMC is remarkably blue relative to other satellites in its magnitude range, in the 0.3 , 1.3 , and 9.9 percentile for the RC3, \citet{eskew11}, and \citet{bothun88parkinglot} colors, respectively (also shown in the legend of Figure \ref{fig:colorhist}).  In the particular case of the RC3 color, the LMC is the bluest object in the sample.  While the other colors are somewhat less extreme, they still place the LMC firmly on the blue side of the blue cloud.  For the more luminous satellites, this placement is partly due to the fact that the red sequence becomes more significant at brighter magnitudes, but even restricted to the magnitude range close to the LMC, the LMC is unusually blue.  Thus, while the LMC/MW system is not particularly unusual in its existence from the perspective of the sizes of the galaxies involved, it is very unusual in the sense that the LMC has an anomalously high star formation rate.

This remarkably large star formation rate for the LMC, especially when compared to similar satellite systems, begs the question of its cause.  The most straightforward explanation for this effect is that the LMC is currently undergoing triggered star formation due to tidal or gasdynamical perturbation by the MW \citep{kenn87,funes04,wgb06,woodsgeller07}.  This is also consistent with star formation histories of the LMC \citep{harris09lmcsfh,masch10lmcsfh} that indicate an increase in the star formation rate in the last 100 Myr - 1 Gyr (although there are some earlier possible peaks).  Thus, the very blue colors of the LMC relative to similar objects are consistent with recent proper motions \citep{kav06lmcpm,piatek08lmcpm} that place the LMC on first infall \citep{besla07lmcfirst,bk10}.  Together  with the result that other luminous satellites are red and all MW satellites aside from the MCs are non-starforming and extremely gas-poor \citep{grcevich09}, the complete picture suggests that this star formation rate cannot be sustained for very long.

Finally, we consider the LG galaxy M33.  If placed in the SDSS, M33 would appear as a secondary of M31 \footnote{M31 would be considered isolated from a distant viewer's perspective unless the MW-M31 separation vector is close to the line of sight, which would occur $\sim 35 \%$ of the time for random orientations.}, with a projected separation of 208 kpc \citep[assuming an M31 distance of 791 kpc from][]{m31dist} and $\Delta v < 500$ $\kms$.  If we take M33 to have $M_V=-18.87$ and $B-V=0.65$ following \citet{vdb00lgbook}, converting to SDSS bands as described above for the MW gives $M_{r,{\rm M33}}=-19.06$ and $g-r=0.44$.  Thus, in a cosmological context, Figure \ref{fig:fractrue} gives an $\sim 86 \%$ probability that M33 is actually a satellite of M31.  However, its $g-r$ color is quite typical for star-forming secondaries (see Figures \ref{fig:matchcmdsec} and \ref{fig:colorhist}), suggesting it does not have the significant triggered star formation evident in the LMC. 

\section{Comparison With Other Work}
\label{sec:comparison}

We now consider these results in the context of recent work examining the frequency of MC-like satellites.  \citet{liu10} examined a similar set of MW/LMC system analogs in the SDSS photometric sample and determined the abundance of satellites.  Our selection is distinguished from these by the use of spectroscopic redshifts to define our clean sample of secondaries, somewhat different selection criteria, and the use of separation distributions instead of a single distance (they adopted 150 kpc).  While our use of the spectroscopic sample reduces our volume relative to that of the main sample of \citet{liu10}, individually, our objects have a much higher likelihood of being true satellites because our redshifts are spectroscopic rather than photometric.   Nonetheless, our results are consistent within the error bars -- the \citet{liu10} result of roughly 20\% probability to host one satellite matches our result at 150 kpc.  We note that, as described in \S \ref{sec:iso}, 250 kpc is more appropriate for asking the specific question of whether or not a host has a satellite within its \emph{virial radius}.

\citet{bk09noplace} and \citet{busha10mwfrommcs} examined exactly this question using high resolution n-body simulations (Millenium II and Bolshoi, respectively).  They both found similar results, finding $\sim 5-30 \%$ of MW analogs host LMCs.  While this may appear at first glance to be at discord with our results, examination of Figure \ref{fig:fracisocosmo} explains this apparent discrepancy.  As the thin solid lines show, the fraction of hosts with satellites within a given separation depends quite strongly on the halo mass.  This fact is taken advantage of by \citet{busha10mwfrommcs} and \citet{bk10} to constrain the properties of the MW based on the existence and observed properties of the MCs.  Further, as illustrated in \citet{busha10mwfrommcs}, the satellite mass range also has a (somewhat weaker) effect on the satellite fraction.  Thus, while the exact definition of both ``MW-like'' and ``LMC-like'' will affect the exact fraction, the effect is generally only a factor of a few for reasonable definitions.  Further, when we test matching definitions, we find identical answers within the relevant error bars.

Additionally, in the context of the color distribution for the SDSS clean sample relative to the LMC, the low numbers of LMC-like satellites reported in \citet{james08ha} and \citet{james10mcs} become clear.  Both are $H\alpha$-selected surveys searching MW-like hosts for satellites with LMC-like R-band magnitudes and $H\alpha$ emission (indicating significant star formation).  Both report LMC-like satellites are rare, implying the LG is a statistical outlier. Figure \ref{fig:colorhist} implies that this apparent anomaly is in fact due to the over-abundance of red secondaries relative to blue for the primaries in question.  \citet[]{james10mcs} Figure 5, in particular, indicates that $8.2 \%$ of hosts have star-forming satellites in the ``LMC'' category within 100 kpc (including their specified volume correction of 1.96).  While the exact selection criteria are slightly different, this result is fully consistent within Poisson error bars of the fraction of hosts with blue cloud galaxies that we find within 100 kpc.  Further, their result that the $H\alpha$-inferred star formation rate of the LMC is higher than most of the other star-forming satellites in  their survey lends further credence to the claim that the LMC has an unusually large star formation rate for its status as a satellite of an isolated $\sim L_*$ host.

\section{Conclusions}
\label{sec:conc}

In this study, we examine a spectroscopic sample of galaxy pairs from the SDSS selected to be composed of an isolated $\sim L_*$ ($M_r < \prilim$, median \medpri) galaxy with a $\sim 0.1 \, L_*$ ($\seclim > M_r > \prilim$, median $M_r=\medsec$) nearest satellite, loosely resembling the MW/LMC system.  We do this by identifying objects with no $\sim L_*$ galaxies nearby (on-sky), and identify fainter galaxies within 250 kpc that have consistent redshifts as likely satellites.  We also examine identically selected halos and subhalos in a mock galaxy catalog from the \MSII{} cosmological simulation.  The high purity of the spectroscopic sample allows us to directly compare secondaries to subhalos, confirming our interpretation of the observations in a \LCDM{} context. It also allows for direct examination of the secondaries' photometric properties.  This is therefore a cosmologically significant sample of ``small-scale structure''.  From this work we draw the following conclusions:

\begin{enumerate}
\item The projected radial distribution (Figure \ref{fig:fracrad}) and line-of-sight velocity distribution (Figure \ref{fig:dvdists}) of bright secondaries matches \LCDM{} predictions.  This implies that there is neither an overabundance nor an underabundance problem for luminous ($M_r \lesssim \seclim$) satellites of isolated $\sim L_*$ ($M_r \sim \medpri$) galaxies.  Specifically, with minimal assumptions (abundance matching) \LCDM{} can reproduce the observed properties of the Universe for galaxies as faint as $L\sim10^9 L_{\odot}$ on velocity scales of $\sim 95 \; \kms$ and spatial scales of $\sim 50$ kpc.

\item Bright secondaries are significantly redder than a general sample of galaxies of the same luminosity (Figures \ref{fig:matchcmdsec} and \ref{fig:colorhist}).  More than half of the $\sim 0.1 \, L_*$  secondaries inhabit the red sequence, compared with only $\sim 30 \%$ in our control sample.  This suggests that environmental quenching is operating within the halos of $\sim L_*$ galaxies, in the absence of harassment from comparably bright satellite counterparts.

\item MW-size galaxies hosting LMC-size satellites are reasonably common in the SDSS sample.  We estimate that $\sim 42 \% $ of $\sim L_*$ galaxies host a $\sim 0.1 \, L_*$ satellite within the virial radius of their dark matter halos.        Further, 10 \% of isolated $\sim L_*$ galaxies have a bright secondary within $50$ kpc, the distance to the LMC.  The exact fraction depends on the mass range used for the host, however, fully accounting for the mild difference between this and other similar measurements.   In any case, this implies that the presence of a satellite with the luminosity of the LMC in a MW-like galaxy dark matter halo is \emph{not} a major anomaly in a cosmological context. 

\item The LMC is remarkable insofar as it is one of the bluest satellites in the sample.  This color is a consequence of an unusually high star formation rate, and studies of the LMC must be cautious when generalizing its star formation properties to other galaxies.  This also suggests the LMC may be undergoing strong triggered star formation if it is on first infall, consistent with the most recent proper motions \citep{kav06lmcpm,piatek08lmcpm}.

\end{enumerate}

In the end, the observations presented here represent another test passed by \LCDM{}, and place the Local Group firmly in the cosmological context of the local universe.  They do, however, beg some questions. First, at what scale, if any, do the standard \LCDM{} abundance matching assumptions break down? This work shows that it is successful for $v_{\rm max} \gtrsim 95 \; \kms$, while the results of \citet{bk11toobigtofail} suggest problems with MW satellites for $v_{\rm max} \sim 50 \; \kms$.  This implies there may be a break down scale somewhere in the range $50 < v_{\rm max} < 95 \; \kms$, but there are no clear causes of such a break down in this range.  An additional important question is that of what causes the quenching of star formation in these secondaries.  Finally, is the remarkable blueness of the Milky Way's LMC just chance timing, or does it imply something more fundamental?  The answers to these questions are well beyond the scope of this work, but it is clear that these isolated host/satellite pairs, while directly comparable to the wider \LCDM{} cosmology, may teach us as much about our own small corner of the Universe.

\section*{Acknowledgements}

The authors would like to thank Julianne Dalcanton, David Nidever, Evan Skillman, Sheila Kannappan, Gurtina Besla, and Coral Wheeler for helpful discussions.  Additionally, we thank Andrew Zentner for substructure models used in early stages of this paper, as well as Michael Eskew and Dennis Zaritsky for kindly providing their LMC photometric history.  We also thank the anonymous referee for clarifying suggestions.  

EJT acknowledges the GAANN Fellowship and UCI Center for Cosmology for support.  MBK is supported by a fellowship from the Southern California Center for Galaxy Evolution.  JSB was supported by NSF grant AST-1009973. CQT acknowledges the generous support of an NSF Graduate Research Fellowship. 

Funding for the Sloan Digital Sky Survey (SDSS) has been provided by the Alfred P. Sloan Foundation, the Participating Institutions, the National Aeronautics and Space Administration, the National Science Foundation, the U.S. Department of Energy, the Japanese Monbukagakusho, and the Max Planck Society. The SDSS Web site is http://www.sdss.org/.

The SDSS is managed by the Astrophysical Research Consortium (ARC) for the Participating Institutions. The Participating Institutions are The University of Chicago, Fermilab, the Institute for Advanced Study, the Japan Participation Group, The Johns Hopkins University, Los Alamos National Laboratory, the Max-Planck-Institute for Astronomy (MPIA), the Max-Planck-Institute for Astrophysics (MPA), New Mexico State University, University of Pittsburgh, Princeton University, the United States Naval Observatory, and the University of Washington.

The Millennium and Millennium-II simulation databases used in this paper and the web application providing online access to them were constructed as part of the activities of the German Astrophysical Virtual Observatory.

{\it Facilities:} \facility{SDSS}

\bibliography{lmcanalogs}{}
\bibliographystyle{hapj} 

\end{document}